\pdfoutput=1 
\documentclass[authoryear,10pt,5p,times]{elsarticle}

\usepackage{hyperref}

\journal{Astronomy and Computing}

\begin{document}
\begin{frontmatter}
\title{Calibration of LOFAR data on the cloud}

\author[ifa,iaa]{Jose Sabater}
\ead{jsm@roe.ac.uk}

\author[iaa]{Susana S{\'a}nchez Exp{\'o}sito}
\ead{sse@iaa.es}

\author[ifa]{Philip Best}
\ead{pnb@roe.ac.uk}

\author[iaa]{Juli{\'a}n Garrido}
\ead{jgarrido@iaa.es}

\author[iaa]{Lourdes Verdes-Montenegro}
\ead{lourdes@iaa.es}

\author[bsc]{Daniele Lezzi}
\ead{daniele.lezzi@bsc.es}

\address[ifa]{Institute for Astronomy (IfA), University of Edinburgh, Royal
Observatory, Blackford Hill, EH9 3HJ Edinburgh, U.K.}
\address[iaa]{Instituto de Astrof\'{\i}sica de Andaluc\'{\i}a, CSIC, Apdo.
3004, 18080, Granada, Spain}
\address[bsc]{Barcelona Supercomputing Center (BSC), Carrer de Jordi Girona, 
29-31, 
08034 Barcelona, Spain}

\begin{abstract}
New scientific instruments are starting to generate an unprecedented amount of 
data. The Low Frequency Array (LOFAR), one of the Square Kilometre Array (SKA) 
pathfinders, is already producing data on a petabyte scale. The calibration of 
these data presents a huge challenge for final users: a) extensive storage and 
computing resources are required; b) the installation and maintenance of the 
software required for the processing is not trivial; and c) the requirements of 
calibration pipelines, which are experimental and under development, are quickly 
evolving. After encountering some limitations in classical infrastructures like 
dedicated clusters, we investigated the viability of cloud infrastructures as a 
solution. 

We found that the installation and operation of LOFAR data calibration pipelines 
is not only possible, but can also be efficient in cloud infrastructures. The 
main advantages were: (1) the ease of software installation and maintenance, and 
the availability of standard APIs and tools, widely used in the industry; this 
reduces the requirement for significant manual intervention, which can have a 
highly negative impact in some infrastructures; (2) the flexibility to adapt the 
infrastructure to the needs of the problem, especially as those demands change 
over time; (3) the on-demand consumption of (shared) resources. We found that a 
critical factor (also in other infrastructures) is the availability of scratch 
storage areas of an appropriate size. We found no significant impediments 
associated with the speed of data transfer, the use of virtualization, the use 
of external block storage, or the memory available (provided a minimum threshold 
is reached). 

Finally, we considered the cost-effectiveness of a commercial cloud like Amazon 
Web Services. While a cloud solution is more expensive than the operation of a 
large, fully-utilised cluster completely dedicated to LOFAR data reduction, we 
found that its costs are competitive if the number of datasets to be analysed is 
not high, or if the costs of maintaining a system capable of calibrating LOFAR 
data become high. Coupled with the advantages discussed above, this suggests 
that a cloud infrastructure may be favourable for many users.
\end{abstract}

\end{frontmatter}

 
\section{Introduction}
\label{intro}

In the 21st century, scientific research is being shaped by the exponential 
increase of the amount of data generated by new scientific instruments. Capture, 
curation, analysis, and sharing of these huge data volumes is a challenge that 
has triggered a new scientific paradigm: data-intensive science \citep[`The 
Fourth Paradigm';][]{Hey2009}. The astronomy community is preparing for the 
forthcoming Square Kilometre Array \citep[SKA;][]{Ekers2012}, an instrument that 
once built, will be the largest scientific infrastructure on Earth and will 
achieve data rates on an exabyte scale. Currently some SKA pathfinders like the 
Low Frequency Array \citep[LOFAR;][]{LOFAR} are already producing data on a 
petabyte scale. These scientific data together with those from the so-called 
`Internet of Things', define the Big Data challenge.

To face this challenge, both powerful computing and high-capacity storage 
resources are required, as well as procedures capable of extracting relevant 
information from the data and sharing it while ensuring reproducibility. 
Algorithms like Map-Reduce \citep[e.g.][]{Lammel2008} have been key to process 
the unstructured data distributed on the Internet. However, they are not 
suitable for some scientific use cases, so scientists need to build procedures 
that are able to both process complex data and efficiently exploit the computing 
resources. A prime example of this is computational genomics, where algorithms 
like BLAST \citep{Altschul1990} have taken advantage of supercomputing resources 
to empower genome sequence searches. The scientific community also investigates 
new computing infrastructures. The Large Hadron Collider (LHC) project designed 
a Grid tiered model, the Worldwide LHC Computing Grid (WLCG), that led to the 
creation of an European Grid Infrastructure open to other scientific 
communities. On the other hand, cloud computing is arising as a more flexible 
model than Grid computing or supercomputing, offering a virtual environment that 
is able to adapt to different use cases.

The LOFAR telescope is characterised by a new design that utilises software 
solutions to implement functionalities, like data correlation or source 
targeting, that have traditionally been performed by hardware cards or 
mechanical devices. The pipelines for reducing LOFAR data are 
high-computational-demand software, and indeed LOFAR capabilities are limited by 
the available computing power instead of by the available observing time as in 
other existing telescopes. Therefore, in order to speed up the data processing, 
it is necessary to research both algorithms and computing resources.

We stress that the study that we present here was motivated by our real needs as 
final users (the users that receive the data from the observatory and attempt to 
exploit it scientifically) to calibrate LOFAR data (see also \citet{Dodson2016} 
for their similarly-motivated investigation into options for reducing and 
analysing spectral line data from the Very Large Array radio telescope). After 
the LOFAR data were successfully observed and delivered to us, it was clear that 
typical calibration strategies would not suffice for several reasons: a) the 
size of the data, with at least a couple of TB per observation, was large enough 
to require dedicated storage; b) the installation and update of the required 
specialized software was not trivial; c) the calibration process was 
experimental and the development of the new calibration strategies required 
frequent changes to the pipeline; and, d) the computational requirements of the 
pipeline were high enough to demand some type of parallelization. The solution 
widely adopted in the LOFAR community for the analysis of these data is the use 
of dedicated local clusters. After exploring the use of these, and also GRID 
infrastructures, we investigated whether a cloud infrastructure would be a 
suitable candidate technology for these analyses \citep{Berriman2012}. We tested 
different cloud infrastructures, and found that a cloud infrastructure that 
fulfils a given set of requirements may offer a solution for the calibration of 
new radio-interferometry data.

This paper presents the results of the tests that we performed on different 
computing resources, in order to evaluate which of them best fulfils the 
requirements from a new strategy for calibrating LOFAR data. In 
Section~\ref{lofar} we describe the LOFAR telescope and the challenge of 
calibrating its data. In Section~\ref{infrastructures} we outline the 
infrastructures tested, and we present the results of the preliminary tests 
performed in dedicated clusters and Grid infrastructures in 
Section~\ref{preliminary}. The description of the tests performed for the cloud 
infrastructures, and the results obtained, are presented in 
Section~\ref{radio_cloud}, including a comparison of costs relative to dedicated 
clusters. We review the suitability of the infrastructures tested in 
Section~\ref{suitability} and, finally, we present the summary, conclusions and 
future work in Section~\ref{conclusions}.

 
\section{The Low Frequency Array, LOFAR}
\label{lofar}

The Low Frequency Array or LOFAR \citep[][http://www.lofar.org/]{LOFAR} is a new 
generation radio interferometer that covers low frequencies from 10 to 240 MHz. 
It is an interferometric array of 50 `stations' of relatively simple, low-cost 
dipole antennas, in which each antenna station has no moving parts, but instead 
operates as a phased array driven in software by powerful station-level 
computing. The core of the array is located in the north of the Netherlands and 
there are several remote stations across Europe. Due to its new design and 
technologies, it is considered to be one of the pathfinders of the Square 
Kilometre Array \citep{Ekers2012}. The output data from each LOFAR station is 
streamed (at up to 3 Gb/sec per station) to a Central Processing facility based 
in Groningen, which handles real-time data operations such as the correlation of 
the data streams. The data output from the correlator is then streamed to an 
offline Central Processing facility, on which application-dependent offline 
processes are run on the data. For standard survey imaging these may involve 
flagging of corrupted data (e.g.\ due to radio-frequency interference) followed 
by some averaging of the data in time and frequency to reduce data volume. There 
is local storage on the central processor to retain datasets for a short period 
of time whilst this processing occurs. The data are then transported to the Long 
Term Archive (LTA), where they are made available to users. The LTA is based on 
a Grid infrastructure \citep{Holties2012}.

Many of the SKA pathfinders are conducting large sky surveys \citep{Norris2013} 
and LOFAR has already presented the first results from a shallow Northern-sky 
survey, the Multi-frequency Snapshot Sky Survey \citep[MSSS;][]{Heald2015}, a 
first step towards the production of deeper imaging surveys. These deeper 
surveys are being performed by the LOFAR Surveys Key Science Projects, which 
aims to explore the low frequency radio sky with an unprecedented depth 
\citep{Rottgering2011}. A component of this is the LOFAR Two-metre Sky Survey 
\citep{Shimwell2016barXiv} which is a wide area survey of the northern sky, 
reaching a typical r.m.s. within each 8-hr observation of $\sim100\,\mu$Jy/beam 
at 150 MHz, substantially deeper than any previous large-area radio survey 
\citep{vanWeeren2016b, Williams2016, Shimwell2016, Hardcastle2016}. Another 
component of the survey is to observe some selected fields, in which the 
highest-quality multi-frequency data exist, in two deeper tiers of observations. 
The work presented here was motivated by the need of calibration and analysis of 
one of the deepest fields: the ELAIS-N1 field (Right Ascension \mbox{16:08:44} 
and Declination \mbox{+56:26:30}). With about 200 hours of observations so far 
and more than 60 TB of pre-processed data in the LTA, the calibration and
analysis of these data presents a formidable challenge.

\subsection{Description of the data}
\label{lofar_data}

In this paper, we focus on the generation of deep wide-field radio images of the 
sky at low frequencies, ignoring all of LOFAR's other observation modes and 
techniques \citep[e.g.][]{Breitling2015}. For our tests, we use an ELAIS-N1 
field dataset, observed using LOFAR's High-Band Antennas (HBA; these cover the 
110 to 240\,MHz frequency range). This observation is broadly typical of any 
LOFAR Surveys dataset: our results are thus generic for the calibration and 
imaging of all LOFAR HBA interferometric data.

Typical LOFAR imaging observation runs consist of: a) a 5-10 minute observation 
of a primary flux calibrator; b) several hours (5 to 8 hours, depending on the 
position of the target field and the data) on the main target, and either some 
flanking fields or a secondary target, using the multi-beam capacities of LOFAR; 
c) a final 5-10 minute run on a primary flux calibrator. The resulting LOFAR UV 
data is stored in CASA Measurements Set format \citep{MS_definition, 
vanDiepen2015}. The field-of-view of the LOFAR HBA is of order 6 degrees (full 
width at half maximum; dependent on the observing frequency), with a beam 
resolution of $\sim 5$ arcseconds.  The frequency coverage for the ELAIS-N1 data 
ranges from 115 MHz to 175 MHz in 371 separate sub-bands. Each sub-band was 
originally composed of 64 spectral channels, and the initial scan-time was set 
to 1 second. During pre-processing, four of the edge-most channels were removed 
(due to poorer sensitivity) and the remaining data were averaged down to 15 
channels per sub-band and a scan time of 2 seconds, reducing the size of the 
dataset by a factor of 8. The typical size of a full pre-processed observation 
is of the order of 3 to 4 TB.

The size of the datasets involved makes necessary a separation
of the data based on different parameters. The calibration process usually
takes advantage of correlations between the data to reduce the degrees of
freedom of the problem and thus the bias of the fitting process
\citep[while not over-fitting; see][]{vanWeeren2016}. Hence, the
parameters that are less correlated are selected to split the data:

\begin{description}
 \item[Time:] The UV data are usually correlated in adjacent time slots but
   there is little correlation in data taken far apart in time. This is
   particularly the case when data are split in different observing runs
   from different days. In this case, the final data model
   (description of the sources of the field) obtained in one observation
   can be used as an input model for another observation. Although the
   LOFAR data used in our tests is split in long chunks of 5 to 8 hours of 
   data, the use of shorter time chunks is currently being tested.
 \item[Sky field:] Different fields on the sky observed simultaneously (using
   multi-beam capabilities) contain different sources, unless they
   overlap. Usually their calibration solutions are substantially
   different (especially at low frequencies) and, in this sense, there is 
   little advantage in processing the fields simultaneously. Therefore, the 
   data are generally split into different sky fields that are processed 
   separately.
 \item[Frequency:] There are correlations between data at different
   frequencies that are affected by the drift of the reference clock signal of 
   the observing stations \citep[e.g.][]{vanWeeren2016} and the effect of the 
   plasma in the ionosphere. However, the structure of the data makes it easy 
   to separate them in frequency ranges or sub-bands. The user can then decide 
   how many of these sub-bands are required for a calibration in order to 
   maximize the performance of the fitting process while keeping the size of 
   the data manageable.
\end{description}

\begin{figure*}
 \centering
 \includegraphics[width=\textwidth]{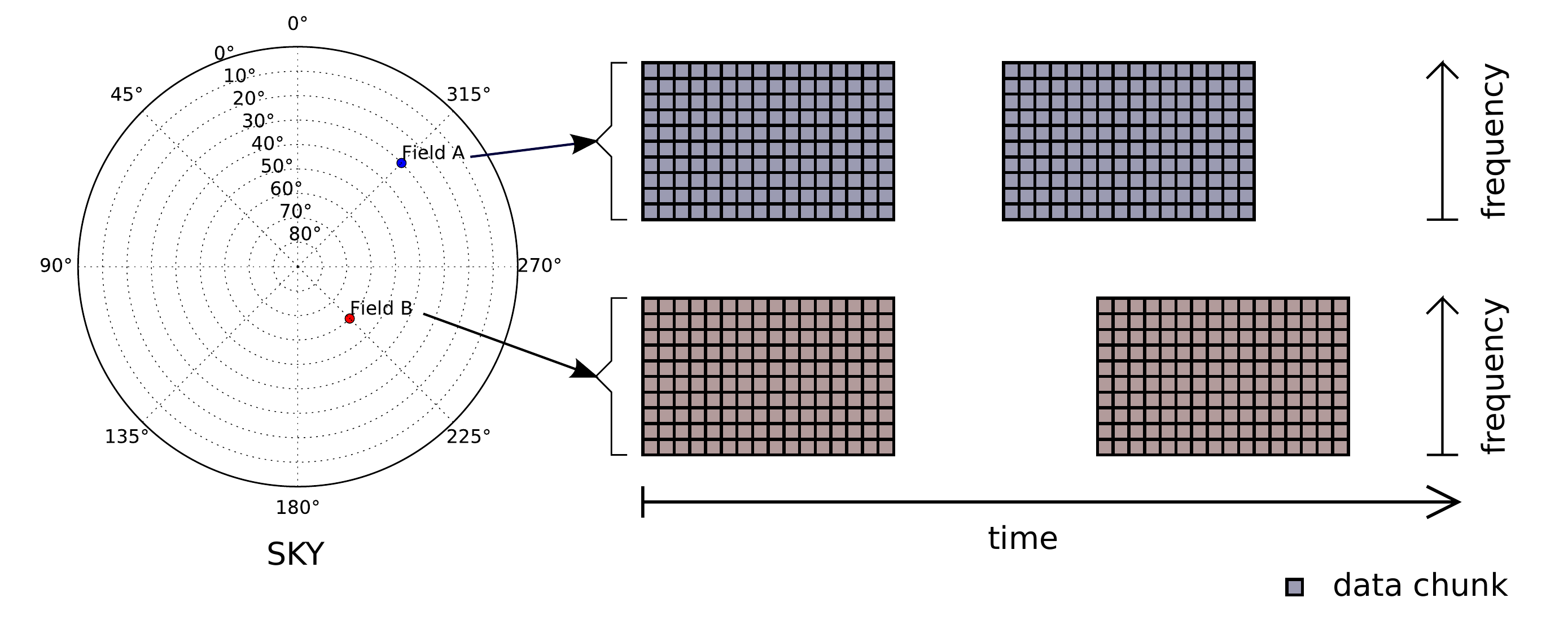}
 \caption{Partition of data in data chunks for parallel processing. They can be 
separated in different sky fields, time blocks and frequency blocks.}
 \label{fig:data_chunks}
\end{figure*}

Fig.~\ref{fig:data_chunks} shows a schema of the partition of data using the 
sky field, time and frequency parameters.

\subsection{Challenges posed by the data calibration}
\label{challenges_data}

LOFAR is currently completely operational and generating high quality data, 
whose correlation and pre-processing is performed routinely by the observatory. 
However, the calibration of the data by the final user still presents many 
challenges.

\subsubsection{Specialized software}
\label{software}

The calibration of the data requires dedicated LOFAR software. CASA 
\citep{McMullin2007} can be used for some general tasks but some specialized 
tasks are only implemented by the LOFAR software compilation. These are mainly 
tasks that require the complex beam model of LOFAR, or that are optimized for 
the use on LOFAR data. The installation of the LOFAR software was especially 
difficult in old systems. It is relevant to take into account that the local 
computing resources available to the final user may be limited to these old 
systems. Although the complexity of the installation process has vastly improved 
due to the efforts of the LOFAR developers and packagers, the final user is 
required to maintain the software up-to-date, in order to avoid critical bugs 
that are corrected in new releases. With several releases each year (e.g. 12 
major releases in 2015 and 2016 plus several individual bug fixes), it may not 
be an easy task to keep the system up to date for a final user.

\subsubsection{High requirements on computing and storage 
resources}
\label{big_data}

A large amount of both storage and computing resources are needed for the 
calibration of these data. The final data set for a single observation run can 
amount up to several TB and a simple calibration run could take several 
CPU-years. However, the pipeline design allows the processing of different 
chunks of data (see Sect.~\ref{lofar_data}) as independent jobs, which 
individually do not require a high amount of memory or processors. This suggests 
that a High Throughput Computing (HTC) system would fit better than a High 
Performance Computing (HPC) one. It is therefore advantageous to use 
powerful distributed computing infrastructures (DCIs) that may or may not be 
readily available to the final user.

\subsubsection{Varying hardware requirements}
\label{ionosphere}

One of the scientific challenges that LOFAR needs to overcome, and that will be 
even more demanding for SKA1 Aperture Arrays, relates to the 
necessity for `direction-dependent calibration', which greatly increases the 
required computation. This arises both from the varying beam that the LOFAR 
stations produce, and the effects of the ionosphere, which are especially strong 
at low frequencies \citep{Intema2009}. In this latter regard, LOFAR poses one of 
the most difficult cases, with both a wide field of view and very long 
baselines, meaning that not only do different stations observe through a 
different patch of ionosphere, but within the imaged field-of-view of each 
station the ionospheric conditions are different towards different sources. The 
experimental pipeline developed by \citet{vanWeeren2016} aims to correct for 
these direction-dependent ionospheric effects. The pipeline has hardware 
requirements that have changed as it is developed. For example, new releases of 
the pipeline tend to require less memory and allow tasks to be run in parallel.  
On the other hand, the increased efficiency has permitted the processing of a 
higher volume of aggregated data in a single computing node, and the overall 
requirements are again increased. An alternative method of applying direction 
dependent calibration is also under investigation (Tasse et al. in prep.), and 
this has yet another different set of hardware needs. The key general point here 
is that during the development process of new analysis techniques, although the 
global efficiency increases, the hardware requirements can often vary 
significantly. Flexibility is therefore a major advantage for the final user.

\section{Computational infrastructures investigated}
\label{infrastructures}

In this section we describe three different kinds of infrastructures in which 
we have performed our tests.

\subsection{Dedicated cluster}
\label{dedicated_cluster}

The most widely adopted solution for the calibration and analysis of LOFAR data 
is the use of a dedicated cluster. We installed the LOFAR software and tested 
the pipeline described in Sect.~\ref{preliminary} locally on the cluster at the 
Instituto de Astrof\'{\i}sica de Andaluc\'{\i}a - Consejo Superior de 
Investicaciones Cient\'{\i}ficas (IAA-CSIC) and in the Stacpolly cluster at the 
Institute for Astronomy of the University of Edinburgh. We also tested the 
pipeline in the LOFAR-UK Computing Facility in the University of Hertfordshire. 
They are typical institutional supported clusters that use a PBS queue system to 
manage the jobs. The IAA-CSIC cluster and the LOFAR-UK Computing Facility use a 
shared file-system that is mounted in all the computer nodes while Stacpolly 
stores the data in a dedicated node.

\subsection{Grid infrastructure}
\label{grid}

A Grid infrastructure was also considered for our tests. A computing Grid 
gathers the storage and computing resources provided by different sites or 
nodes, that can be seen as clusters connected through a Wide Area Network (WAN) 
running a common middleware that allows them to work as a single infrastructure. 
The access and the usage of the Grid resources are managed by means of Virtual 
Organisations (VOs). Users are grouped in VOs, and Grid nodes, depending on 
their institutional policy, support one or more VOs. Grid infrastructures 
provide a distributed parallel execution environment which is ideal to process 
this kind of data.

\subsection{Cloud infrastructure}
\label{cloud}

Finally, we considered a cloud solution which can offer the flexibility 
required by the target problem. 

A cloud infrastructure provides shared computing, networking, data storage, or 
other services, on-demand in a flexible way. Most infrastructures offer the 
processing resources as computing instances (mainly virtual machines based on a 
configurable template or ``image'') that can be quickly provisioned or released. 
They can also offer object storage, in which the data are viewed as objects in a 
hierarchy, or block storage, in which data are treated as blocks of bytes like 
in classical disks and file-systems.

The tests were performed on three different cloud systems: a) an academic 
private multi-site cloud system, the European Grid Infrastructure (EGI) 
Federated Cloud, that brings together computing and storage resources from 
different national and intergovernmental European providers; b) an academic 
private one-site cloud managed by the Rutherford Appleton Laboratory (RAL), 
provided by the Science \& Technology Facilities Council (STFC) in the United 
Kingdom; and, c) a commercial multi-purpose public multi-site cloud, the Amazon 
Web Services (AWS) infrastructure.

The EGI Federated Cloud is a private cloud infrastructure open to any research 
community. It has been developed in the framework of EGI.eu and currently 
federates about 20 sites whose computing and storage resources are owned by 
different academic European institutions. In this infrastructure, computing and 
storage resources are provided through the standard Open Cloud Computing 
Interface (\href{http://occi-wg.org/}{OCCI}) and Cloud Data Management Interface 
\citep[CDMI;][]{CDMI} and users access them with their X.509 certificates as 
member of a specific Virtual Organisation (VO). Once the user gets their proxy 
certificate, they can use it to gain access to the resources either through a 
Ruby or Java OCCI client or through high level tools built upon OCCI connectors 
that allow the user to manage the virtual resources. One example is COMPSs 
\citep{Lordan2014}, a programming model that, apart from being able to act as a 
cloud broker, optimizes the use of the computing resources through the 
exploitation of the inherent parallelism of the scientific applications.

Both the EGI Federated Cloud and the LOFAR LTA share a common Authentication and 
Authorization (A\&A) method based on a Public Key Infrastructure that uses the 
X.509 standard. This framework is defined by the European Policy Management 
Authority for Grid Authentication in e-Science Certification Authorities 
federation (\href{https://www.eugridpma.org/}{EUGridPMA}), and 
the Interoperable Global Trust Federation 
(\href{https://www.igtf.net/}{IGTF}). After obtaining their personal 
certificate, the users need to install the voms-client tool and to properly 
configure it. Users get a valid certificate proxy through this tool and can 
issue OCCI commands to the EGI Federated Cloud or request the transfer of data 
from the LOFAR LTA.

The Rutherford Appleton Laboratory Cloud offers another private cloud 
infrastructure open to the British scientific community. It is based on 
\href{http://opennebula.org/}{OpenNebula} and provides mainly computing 
resources. The access to the infrastructure is provided by a custom dashboard 
and their installation of the OpenNebula Cloud operations centre called 
Sunstone. It is also possible to leverage the OpenNebula Application Programming 
Interface (API) endpoint using XML-RPC. Recently, a 
\href{http://ceph.com/}{Ceph} object store has been commissioned and integrated. 
Ceph offers an AWS-compatible API which simplifies the interaction with the 
system.

The third cloud system investigated was Amazon Web Services which is a suite of 
cloud-computing services provided by Amazon Inc. The resources are distributed 
in several regions (13 at the time of writing) around the world. We used the 
biggest one that is located in North Virginia (or ``us-east-1'') to maximize the 
number of resources available. Among the services provided, we used Amazon 
Elastic Compute Cloud (EC2), Amazon Simple Storage Service (S3) and Amazon Route 
53. EC2 is the Infrastructure as a Service part, providing the computing and 
block storage resources (Elastic Block Storage or EBS). S3 is the object storage 
service and Route 53 provides managed Domain Name System services. We interacted 
with the services using the web-based AWS console and the AWS API through the 
Python \href{https://github.com/boto/boto}{Boto} library or their 
\href{https://github.com/aws/aws-cli}{command line interface} software.

The SKA Organization published in 2015 a joint call for proposals with Amazon 
(Astrocompute) to investigate and develop radio-astronomy tools and processes in 
the AWS infrastructure. Most of the AWS tests shown here were carried within the 
context of one Astrocompute project proposed by the authors and consumed the 
credits (equivalent to money that can be spent on AWS resources) provided for 
the project. The usage of resources provided by AWS is billed monthly and the 
amount depends on the type of resource and its usage.

\section{Preliminary tests in dedicated clusters and the Grid}
\label{preliminary}

For most of the tests presented in this paper we used a simplified pipeline in 
which the data from just one sub-band are calibrated, corrected and imaged. 
These are the basic steps of a typical self-calibration loop and contain the 
fundamental steps used in calibration pipelines of radio-interferometric data. 
The calibration and correction are run using the LOFAR Black Board Selfcal (BBS) 
system \citep{BBS} and the imaging is done with the AWimager \citep{Tasse2013}. 
BBS runs in one core at the time of making the tests. The data are loaded in 
memory in chunks whose size is manually selected using an input parameter. 
AWimager has multi-core and automatic memory managing capabilities. Although the 
advanced pipelines presented in \citet{vanWeeren2016} are also being adapted to 
the cloud by the authors, the tests using the simplified pipeline cover the 
basic performance issues that can be found in the advanced ones. 

In order to check the general capability of the infrastructure to run the 
pipeline in parallel, we ran the pipeline on several instances in parallel as a 
test. The calibration of a full dataset would require us to run in parallel 
chunks of data obtained from the combination of several sub-bands. However, the 
test on those instances was enough to check how the infrastructure responded to 
the parallel run of the pipeline, while retaining the option to manually check 
possible errors and reduce the time of the individual tests. All of the 
infrastructures involved responded as expected to the parallelization and the 
calibration time using several instances took practically the same time as the 
calibration of a single band in one instance (as long as there were enough 
resources available).

In this section we describe our tests in two different kind of HTC systems: 
clusters and Grids. The results of the test performed in the cloud 
infrastructures will be present in detail in the next section.

\subsection{Dedicated cluster}
\label{tests_cluster}

One of the main issues that drove us to consider other solutions was the 
difficulty to install and keep updated the required software on the clusters.  
The installation on the IAA-CSIC cluster was done first and took several weeks. 
The long time required for the installation was mainly caused by the relatively 
old version of the operating system in comparison with the libraries required by 
the LOFAR software. Almost all of the dependencies had to be manually compiled, 
installed and configured as a local user. The installation on Stacpolly took 
less time due to the experience acquired in the other cluster, the relatively 
new operating system, and the release by third parties of scripts that helped 
with the installation of the dependencies. However, the software became quickly 
outdated and some new dependencies appeared. It had to be manually recompiled 
which was time consuming.

Although the cluster solution works at present, it may not scale well in the 
future when the load of computing and storage requirements increase. 
Furthermore, at the time of writing, the requirements of the newest advanced 
pipelines are higher than the capacity of the nodes of the the IAA and the 
Stacpolly clusters; this highlights a potential difficulty of the inflexibility 
of dedicated clusters during the development phase of new analytic techniques.

\subsection{Grid infrastructure}
\label{tests_grid}

The problems with the installation and update of the software were similar to 
those shown for the clusters with an additional difficulty arising from the use 
of different operating systems in each site. Additionally, the installation and 
deployment of the user application may require the intervention of the system 
administrators of each provider of the Grid infrastructure where the pipeline 
would run. This may be impractical whilst the calibration pipeline is under 
active development and frequently changing. However, this kind of solution could 
be considered in the future to implement a stable calibration pipeline. This 
scenario would be more favourable if the deployment of the software is eased by, 
for example, the use of containers \citep[see 
\href{https://www.docker.com/}{Docker; }][]{Boettiger2015}.

Once the pipelines are stable, it will be possible to run them using the 
computational resources provided by the same Grid site in which the data are 
stored. This would reduce significantly the overhead due to the transfer of 
data. This approach is currently being explored by the LOFAR Surveys Key Science 
Project team on the \href{https://userinfo.surfsara.nl/}{SURFsara} site.

 
\section{Calibration of radio-astronomy data in the cloud}
\label{radio_cloud}

We performed several tests to assess the suitability and performance of
the cloud platforms used to solve our problem. We checked the
following points:

\begin{description}
 \item[Installation and update of the software:] We tested how easy it was
   to install and manage the required software in the infrastructure. We
   also considered how easy it was to keep the software up to date and the
   level of human intervention and support required.
 \item[Suitability of the platform:] We investigated the suitability of the
   platform to run the pipeline and to be adapted to it. We first checked
   if the platform was capable of providing the resources required by the
   pipeline. We also checked the ability to adapt it easily to new
   requirements of the pipeline.
 \item[Data transfer:] We transferred a high volume of data from the LOFAR LTA 
   to some of the infrastructures. We determined the speed of the transfer and 
   the level of human intervention required.
 \item[Processing performance:] We studied whether it was possible to run the 
   pipeline efficiently in parallel in the cloud distributed environment.  We 
   tested several parameters like the running time, memory consumed, load of 
   the processors, etc.
 \item[Cost:] Finally, we investigated the cost of the commercial cloud that we 
   used, AWS, and compared it to the cost of other infrastructures. 
\end{description}

\subsection{Software installation and update}
\label{software}

The installation and update of the software in the infrastructure was one of the 
most time consuming processes. As mentioned in the previous section, we had 
previously tried the manual compilation and installation of the LOFAR 
dependencies by trial and error in a relatively old Red Hat operating system 
used by one of the clusters. This experience was crucial to consider a cloud 
infrastructure as a good alternative. A cloud infrastructure allows to select 
one base operating system that is convenient for the installation of the LOFAR 
software. The software is installed once in an image that is used as the base 
for all the computing instances.

For the installation of the LOFAR software in the EGI infrastructure we used the 
same approach that we followed for the cluster in Edinburgh. We run an automatic 
installation script provided by \href{http://www.astro-wise.org/}{Astro Wise}. 
The script downloaded fixed versions of the dependencies from the Astro Wise 
server that are known to work. It required some manual tweaking and the manual 
installation of some additional dependencies to adapt it to our needs. Although 
the approach is powerful, as it is independent of the Linux flavour and provides 
compiled software that is optimized for the target operating system, there are 
some caveats: a) the version of the dependencies is fixed; b) there is no 
automatic way to upgrade the system of dependencies; c) the lack of a custom 
required dependency that is not implemented in the system is usually detected at 
run-time; and, d) it could take a considerable time to compile all the 
dependencies. Additionally, the maintenance of a generic script like this can be 
cumbersome and, apparently, it is no longer maintained.

Overall, we followed the recommended method to install user applications on the 
EGI Federated Cloud. We created, by means of an OCCI client, a VM using a 
Virtual Appliance published in the EGI Federated Cloud Applications Database 
formerly known as the Application Database Cloud Market (AppDB; 
\href{https://appdb.egi.eu/}{https://appdb.egi.eu/}). Later, the software was 
installed accessing the VM through SSH public/private key pair. Once the user 
application (the LOFAR and CASA software in our case) was installed, we created 
a VM image from a snapshot of the VM (this process could only be done by site 
administrators at time of performing these tests). After that, we uploaded the 
image to the AppDB. Finally, we requested the inclusion of the image in the VO 
image list so it could be deployed on the Federated Cloud sites supporting the 
VO.

For the RAL and AWS cloud infrastructures we used installation and provisioning 
recipes created in Ansible. \href{https://github.com/ansible/ansible}{Ansible} 
is an automation tool used for application deployment, configuration management 
and orchestration. It can be interfaced using ``recipes'' written in 
\href{http://yaml.org/}{YAML} which make them easy to read and write by human 
beings. The recipes are divided by functionality to allow a granular 
installation of software and an easy upgrade\footnote{The recipes are released 
as \textit{playbook roles} in 
\href{https://github.com/nudomarinero/Astrocompute-ELAIS-N1}{ 
https://github.com/nudomarinero/Astrocompute-ELAIS-N1}}. Ansible permits us to 
use the same recipes in different systems like the RAL federated cloud based on 
OpenNebula and the AWS infrastructure. \footnote{We did not use Ansible in EGI 
because the installation process that we followed on EGI required manual 
tweaking and we did it only once.  However, we can not think of a technical 
reason to not to be able to use Ansible in EGI.}

We used the Long Term Releases of Ubuntu (12.04 and 14.04) as our base operating 
system. At first we created our own Ubuntu Personal Package Archives (PPA) with 
some custom changes on the radio astronomy packages created by 
\href{https://github.com/gijzelaerr}{Gijs Moolenar}. Later, the relevant changes 
were implemented in the main 
\href{https://github.com/radio-astro/packaging}{``radio-astro''} 
\href{https://launchpad.net/~radio-astro/+archive/ubuntu/main}{PPA}. This PPA 
was used as the base for the LOFAR software dependencies. However, the main 
LOFAR package is still compiled separately to allow a quicker response to bug 
patches and software upgrades. Please note that the radio-astro PPA is being 
superseded by the new \href{http://kernsuite.info/}{Kern Suite}.

The previously mentioned effort to package the dependencies of LOFAR (mainly 
\href{https://github.com/casacore/casacore/}{casacore}, 
\href{https://github.com/casacore/casarest/}{casarest} and pyrap or 
\href{https://github.com/casacore/python-casacore}{python-casacore}) in an 
Ubuntu PPA allows the integration of the software in the standard packaging 
system with all its advantages (automatic processing of dependencies, easy 
installation and removal of files, etc.) -- though also with some of its 
disadvantages, like the possible lack of compiler optimization.

Regarding the creation of the base image or template for the instances, for the 
RAL cloud the software was installed in an instance running a base 
Ubuntu system. A snapshot of the instance was created and later published as a 
base image. This step required some support from the RAL. For AWS we used 
\href{https://github.com/mitchellh/packer}{Packer}\footnote{``Packer is a tool 
for creating machine and container images for multiple platforms from a single 
source configuration''; \href{https://www.packer.io/}{https://www.packer.io/}} 
to create an Amazon Machine Image (AMI) automatically. In this case no human 
intervention was required at all.

\begin{table*}
\centering
\begin{minipage}{\textwidth}
\centering
\caption{Software bundling and base image preparation in each infrastructure.}
\label{table:softinfra}
\begin{tabular}{lll}
\hline
Infrastructure & installation of software & creation of base image \\
\hline
\hline
cluster Granada & manual installation & - \\
cluster Edinburgh & installation script & - \\
EGI Federated cloud & installation script & human intervention \\
RAL cloud & PPA+custom LOFAR package & Semi-automated with Ansible and human 
intervention\\
AWS cloud & PPA+custom LOFAR package & Automated with Packer and Ansible\\
\hline
\end{tabular}
\end{minipage}
\end{table*}

Table~\ref{table:softinfra} shows a summary of the installation of software and 
the creation of the base image in different infrastructures.

It should be noted that the level of human intervention required can have an 
impact on the usability of the infrastructure. If the step requiring human
intervention has to be repeated often then the time consumed by this can easily
add up to a considerable amount, which makes it impractical. In our case
the software has to be updated often and the level of human intervention
had a considerable impact on the time required to implement our tests.

\subsection{Suitability of the infrastructures}
\label{suitability}

The calibration pipeline is under development and its requirements change. At 
the time of writing this paper, it needs several tens of GB of memory and a 
local scratch data area of at least 1 TB that is used to hold the intermediate 
data. It benefits from multiprocessing in most of the steps. We needed 
infrastructures that can adapt to those requirements.

A wide range of machine sizes helps in the development of a pipeline that 
optimises the usage of resources. A machine with a large number of CPU cores can 
speed up the execution of tasks that are adapted to multiprocessing but its 
resources can remain largely idle if a single core application needs to be run. 
From a user point of view it is sometimes very difficult or even impossible to 
modify or adapt the underlying software to optimize the usage of resources. In 
this context it makes sense to utilise a cloud infrastructure in which the sizes 
of the machines can be adapted to the optimization level of the software 
available. The capacities of relevant instance types in AWS, in the EGI 
Federated Cloud used for the tests (in our case we used 
\href{https://www.cesnet.cz/}{CESNET}), and in the RAL cloud, are summarized in 
Table~\ref{table:infra}. It is important to note that the CPUs of the AWS 
instances are actually virtual CPUs that made use of the Intel Hyper-threading 
technology \citep{Marr2002}, that means that their performance is roughly half 
of the performance of a physical CPU. This correction factor will be taken into 
account for the comparisons.

\begin{table*}
\centering
\begin{minipage}{\textwidth}
\centering
\caption{Type of instances used for the tests in each cloud infrastructure.}
\label{table:infra}
\begin{tabular}{lllll}
\hline
 instance type & cores & memory & type of storage \\
  & (N) & (GB) &  &  \\
\hline
\hline
\multicolumn{5}{c}{AWS} \\
\hline
 m4.large & 2$^{*}$ & 8 & EBS (450 MB/s) \\
 m4.xlarge & 4$^{*}$ & 16 & EBS (750 MB/s) \\
 m4.2xlarge & 8$^{*}$ & 32 & EBS (1000 MB/s) \\
 m4.4xlarge & 16$^{*}$ & 64 & EBS (2000 MB/s) \\
 m4.10xlarge & 36$^{*}$ & 160 & EBS (4000 MB/s) \\
 c4.xlarge & 4$^{*}$ & 7.5 & EBS (750 MB/s) \\
 c4.2xlarge & 8$^{*}$ & 15 & EBS (1000 MB/s) \\
 c4.4xlarge & 16$^{*}$ & 30 & EBS (2000 MB/s) \\
 c4.8xlarge & 36$^{*}$ & 60 & EBS (4000 MB/s) \\
 m3.large & 2$^{*}$ & 7.5 & EBS and $1\times32$ GB int. \\
 m3.xlarge & 4$^{*}$ & 15 & EBS and $2\times40$ GB int. \\
 m3.2xlarge & 8$^{*}$ & 30 & EBS and $2\times80$ GB int. \\
 c3.xlarge & 4$^{*}$ & 7.5 & EBS and $2\times40$ GB int. \\
 c3.2xlarge & 8$^{*}$ & 15 & EBS and $2\times80$ GB int. \\
 c3.4xlarge & 16$^{*}$ & 30 & EBS and $2\times160$ GB int. \\
 c3.8xlarge & 36$^{*}$ & 60 & EBS and $2\times320$ GB int. \\
 r3.large & 2$^{*}$ & 15.25 & EBS and $1\times32$ GB int. \\
 r3.xlarge & 4$^{*}$ & 30.5 & EBS and $1\times80$ GB int. \\
\hline
\multicolumn{5}{c}{EGI Federated Cloud 
    (\href{https://www.cesnet.cz/}{CESNET})} \\
\hline
 mem\textunderscore medium & 2 & 8 & NFS \\
 extra\textunderscore large & 8 & 8 & NFS \\
 mem\textunderscore extra\textunderscore large & 8 & 32 & NFS \\
 mammoth & 16 & 32 & NFS \\
\hline
\multicolumn{5}{c}{RAL Cloud} \\
\hline
Lofar-Ubuntu-14-8c-32gb-2TB & 8 & 32 & Internal 2TB \\
\hline
\end{tabular}
\begin{list}{}{}
 \item $^{*}$ Virtual hyperthreaded cores.
\end{list}
\end{minipage}
\end{table*}

The amount of memory required for the pipeline, particularly for the imaging 
process, could sum to dozens of GB. In the two clusters, one of the technical 
constraints was the limited and fixed amount of memory available on each node. 
In some cases, the computing nodes did not have enough memory to perform the 
imaging step at an adequate resolution. Most of the cloud infrastructures 
provided a wider range of choices for the node sizes including instances with 
large memory. We needed the possibility to launch VMs with this high amount of 
memory.

One particularity of AWS is the availability of non-reserved instances with 
variable prices that follow the market demand, called \textit{spot instances}. 
Usually their prices are several times lower (they range from $\approx10$ times 
lower to being even 10 times more expensive) than those of standard reserved 
instances and are ideal to run processes that do not require real time 
responsiveness like in our case. A maximum bidding price is set when an instance 
is requested and the instance runs until the market price rises over this limit. 
In this event the instance is automatically shut down. When using spot instances 
the pipeline must be adapted to allow it to be resumed in the event of an 
instance shut-down. In our case we used spot instances whenever it was possible. 
It is important to note that the probability of a shut-down of the spot instance 
anti-correlates with the maximum bidding price; a relatively high price will 
allow the instance to run uninterrupted for a longer time. Hence, longer 
processing times usually require higher bidding prices to avoid interruptions. 
We will discuss the final price per instance that we empirically calculated in 
Section~\ref{costs}.

There were several options for the storage, each with advantages and
disadvantages:
\begin{description}
 \item[Shared file-system:] One option used in one of the clusters was the use 
   of a shared file-system (\href{https://www.gluster.org/}{GlusterFS}). 
   Several dedicated nodes provided access to the storage area which was 
   visible from all the computing nodes. The I/O rate could be limited by the 
   local network connection which could be an issue depending on the problem. 
   However, the simplicity of use of these systems could compensate the 
   possible loss of performance during I/O operations. This option is 
   currently available in AWS (Amazon Elastic File System) but was not used 
   for our tests.
 \item[NFS mount:] Another simple option was the use of an NFS \citep[Network 
   File System;][]{Sandberg1986} mount. This
   configuration was used in one of the EGI Federated Cloud
   sites. Although it worked eventually, this solution presented several
   disadvantages: a) the set up of the system required support from the
   site administrators; b) its configuration and set up was not accessible
   by the API; and, c) it was a custom solution that could not be extended
   to other sites and required the development of site-specific steps on
   the testing pipeline.
 \item[Object storage:] Object storage is used to store large data files
   and it is ideal to store the input data, intermediate data objects
   between the main calibration blocks in large pipelines, and final
   output data. Although this is very useful to store large volumes of
   data for extended periods, the files can not be handled as regular
   files in a file-system by the software.
 \item[Block storage:] Block storage can be attached as data volumes to the
   VM instances. They can be used as scratch areas that are resilient to
   instance shut-downs. They can be used to locally store the intermediate
   data as they are viewed as normal disk by the operating system. Block
   storage is usually offered by cloud platforms as one of their services
   but this is not always the case.  Although this type of storage is very
   convenient, the I/O rate could be limited depending on the
   configuration (see Sect.~\ref{performance}) and it is usually more
   expensive than object storage. Hence, a mixed storage system is usually
   used with block storage.
 \item[Internal storage:] Internal storage is provided in some clouds with 
   some types of instances and is equivalent to a typical scratch data area. 
   This type of storage is ideal to locally store the intermediate data, and  
   offers the best performance in terms of I/O speed. However, such storage is  
   usually ephemeral and, in general, the data is wiped once the instance is 
   shut down. This can be a problem if the instance can be shut down before the 
   full completion of the calibration pipeline, as in the case of spot 
   instances in AWS. 
\end{description}

One of the main problems that we did not foresee until testing the pipeline on 
different infrastructures was the effect of limited availability of scratch data 
areas (internal or block storage). The data required for a calibration run must 
be present in the instance in an standard file-system. Even after splitting the 
data in fields, observation runs and frequency bands, the size of these datasets 
ranges from several hundreds of GB to a few TB. Additionally, some space for the 
final data products and for intermediate-step data is also required. Hence, a VM 
instance requires an standard file-system of the order of TBs attached. The lack 
of internal or block storage, or some limits on the size of the block storage 
devices at the time of performing the tests, prevented us from finishing the 
implementation of the full calibration pipeline in all platforms except AWS.

One of the points that had a big impact on the usability of a platform was its 
maturity. More mature platforms usually present more options and a solid 
ecosystem of tools to leverage them. For example, the usage of the libraries and 
APIs of Open Stack or AWS was very easy. Although these points are not purely 
technical, they had an impact on the time that we spent setting up the tests and 
solving the problems that we encountered.

\subsection{Data transfer}
\label{data_transfer}

We checked the data transfer to two cloud infrastructures that offered
object storage. We used S3 in AWS and the new Ceph cluster at the RAL.

The transfer of the data from the LOFAR LTA requires the use of common A\&A 
X.509 credentials as explained in Sect.~\ref{cloud}. The software used for the 
transfer was the \href{https://www.dcache.org/downloads/1.9/srm/}{``SRM'' suite} 
and is based on the \href{https://sdm.lbl.gov/srm-wg/doc/SRM.v2.2.html}{Storage 
Resource Manager protocol}. One dataset (a single observation run ranging from 5 
to 10 hours) was transferred at a time and the process was the following:
\begin{enumerate}
 \item A proxy certificate for the Grid credentials with the longest valid
   time (one week) was created.
 \item A dataset, composed of several data files (usually from 244 to 371
   files), was staged in the LOFAR LTA servers using a XML-RPC petition. During 
   this process the data was moved from tapes to a temporary disk storage area 
   (the staging area) that can be directly accessed for a period of 
   time.
 \item The data was downloaded in parallel from one VM instance in the
   target infrastructure. As soon as a data file was downloaded it was
   uploaded to the object storage service provided by the infrastructure. For 
   this process we wrote some simple
   \href{https://github.com/nudomarinero/AWS_elais-n1_public_data}{Python
     scripts}.
\end{enumerate}

The AWS S3 and Ceph cluster each provide a compatible API. The API can be 
accessed using the Python library Boto in the same way for both infrastructures. 
The only difference was a small set of configuration parameters. This 
compatibility made it very easy to write common software for the upload of data 
that could be used in both infrastructures.

We transferred 18 full datasets from the LOFAR LTA to the AWS S3 facilities in 
North Virginia (us-east-1 region). The volume of each individual dataset is 
about 3 to 4 TB with a total size of $\approx$ 64 TB. An instance of type 
\textit{m4.10xlarge} was launched and provisioned using Ansible. This type of 
instance offered the highest available bandwidth and a large number of 
processors to allow a parallel download. We attached an EBS volume as a 
temporary data area that was used to store the partial data files in the process 
between the start of the download and the end of the upload to the object 
storage service. After testing a number of parallel downloads between 18 and 36 
(36 was the number of processors of the instance), we selected 24 parallel 
downloads as the optimal number that maximized the transfer speed whilst 
maintaining the stability of the instance. We think that the stability problems 
with high concurrencies originated from the saturation of the network connection 
which is also used to communicate to the EBS filesystem.

\begin{figure}
 \centering
 \includegraphics[width=7.5cm]{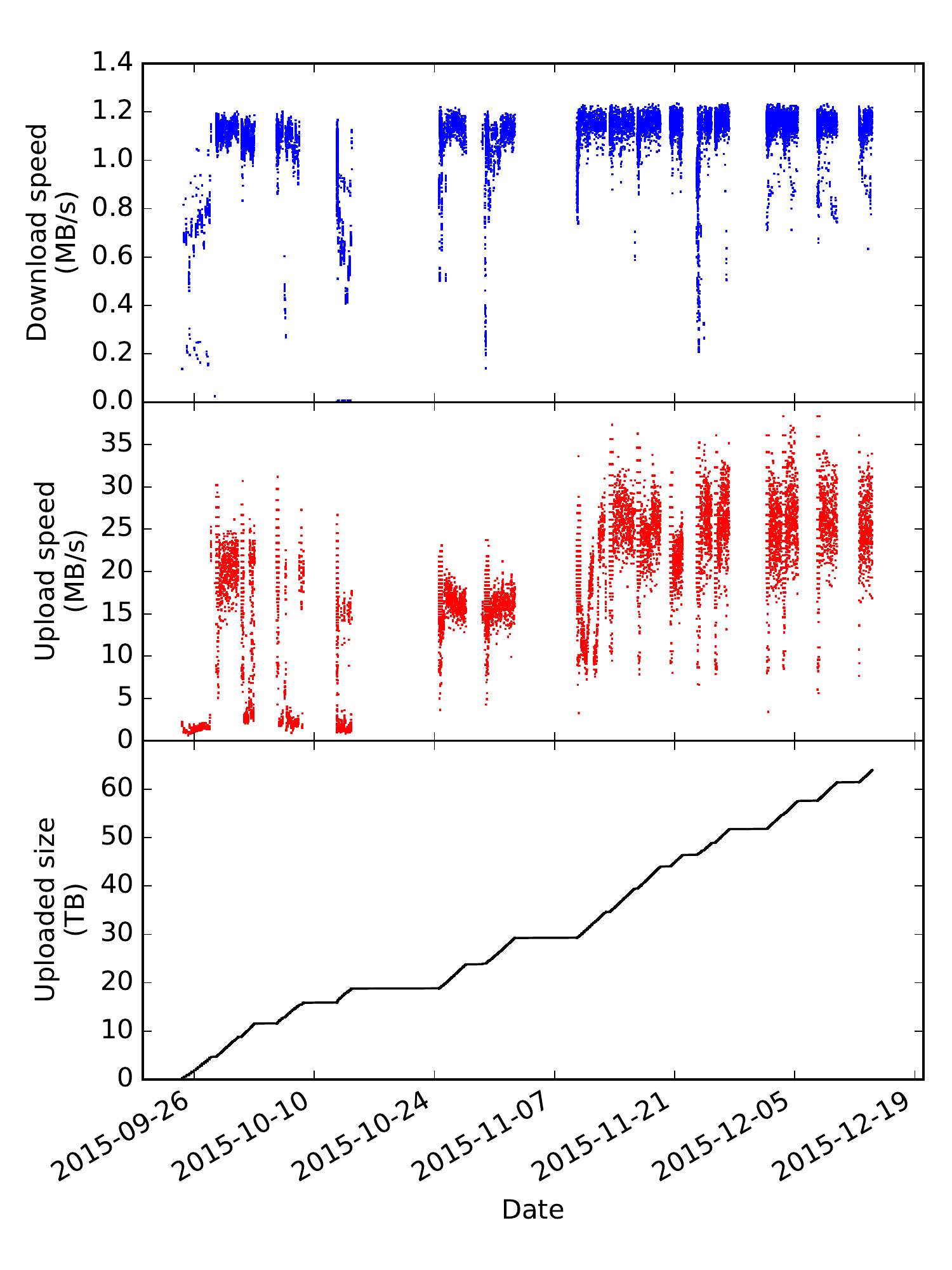}
 \caption{Data transfer to AWS S3. The upper panel shows the speed of transfer 
for each of the data files from the LOFAR LTA to the AWS EC2 instance in the 
us-east-1 region. The mid panel shows the speed of transfer of the data 
files from the AWS EC2 instance to AWS S3. The lower panel shows 
the cumulative size of the data uploaded to S3.}
 \label{fig:transfer}
\end{figure}

The transfer started mid-September 2015 and finished mid-December 2015 as shown 
in Fig.~\ref{fig:transfer}. There are several gaps in the transfer process and 
removing them the total transfer time could be reduced from $\approx$ 3 months 
to $\approx$ 1.5 months. However, most of these gaps were very difficult to 
avoid and probably present a realistic scenario to be found in semi-supervised 
data transfer processes carried out by final users like in this case. The 
transfer speed for each file was measured by dividing its transfer time by its 
size. The transfer speed from the LOFAR LTA to the AWS infrastructure peaks at 
about 1.2 MB/s for each data file. The mean upload speed to S3 (that uses the 
internal AWS network) ranges from less than 15 to more than 35 MB/s for each 
data file. This variance is probably affected by the number of parallel uploads 
and we estimate that the total upload speed at a given point is very similar.

The gaps in the data transfer had different origins. Apart from the testing 
nature of this experiment that made us to spend some time in tests that are not 
shown in Fig.~\ref{fig:transfer}, we had some issues with the staging of data 
and the manual renewal of the proxy certificate. The time required for 
the staging of data in the LOFAR LTA varied widely from time to time, probably 
due to the load on the service. On one occasion the data disappeared from 
the staging area. This was probably caused by the limited storage space of 
the staging area, an issue that is currently being solved by the LOFAR LTA. 
Hence, all the staging process had to be supervised to spot and solve possible 
problems. Additionally, the download from the LTA required the 
creation and usage of a proxy certificate. This process could only be done 
interactively and the maximum duration of the proxy certificate was one week. 
Therefore, the process required constant supervision.

We also uploaded one of the datasets to the Ceph cluster at the RAL. The process 
was similar to the upload to AWS. We used an internal VM instance that was 
provisioned with Ansible and an attached scratch data volume in the RAL cloud. 
Although we could not save the exact profiling data for this test, the transfer 
speed from the LOFAR LTA to the RAL infrastructure was several times (about 3 or 
4 times) higher than the transfer to the AWS infrastructure. We assumed that 
this was mainly caused by a higher bandwidth of the connection between the two 
European sites.

Finally, we would like to mention a possible alternative method to transfer 
data: the physical transfer of storage units (disks, tapes, etc) between data 
centres. After we started our tests, AWS started offering a new service of data 
transfer called Snowball. A unit of 50 TB of solid disk storage with high 
bandwidth 
local network connectivity could be sent by courier to transfer the data between 
the data centre and AWS S3. At the time of writing this paper, the physical 
transfer of data could outperform the transfer via the regular Internet network.

\subsection{Processing performance}
\label{performance}

We focused our performance tests on the EGI Federated Cloud Infrastructure and 
on the AWS infrastructure.

We ran the simplified test pipeline using one single ELAIS-N1 sub-band centred 
at 115.037537 MHz with one frequency channel, 61 antennas (1891 baselines) and 
2879 time slots in an 8 hours observing run (10 s scan time). The size of the 
data was $\approx 1$ GB. The pipeline was run in several instance types in AWS 
and the EGI Federated cloud (see Table~\ref{table:infra}). We measured the time 
spent in each phase of the calibration pipeline.

\begin{figure}
 \centering
 \includegraphics[width=7.5cm]{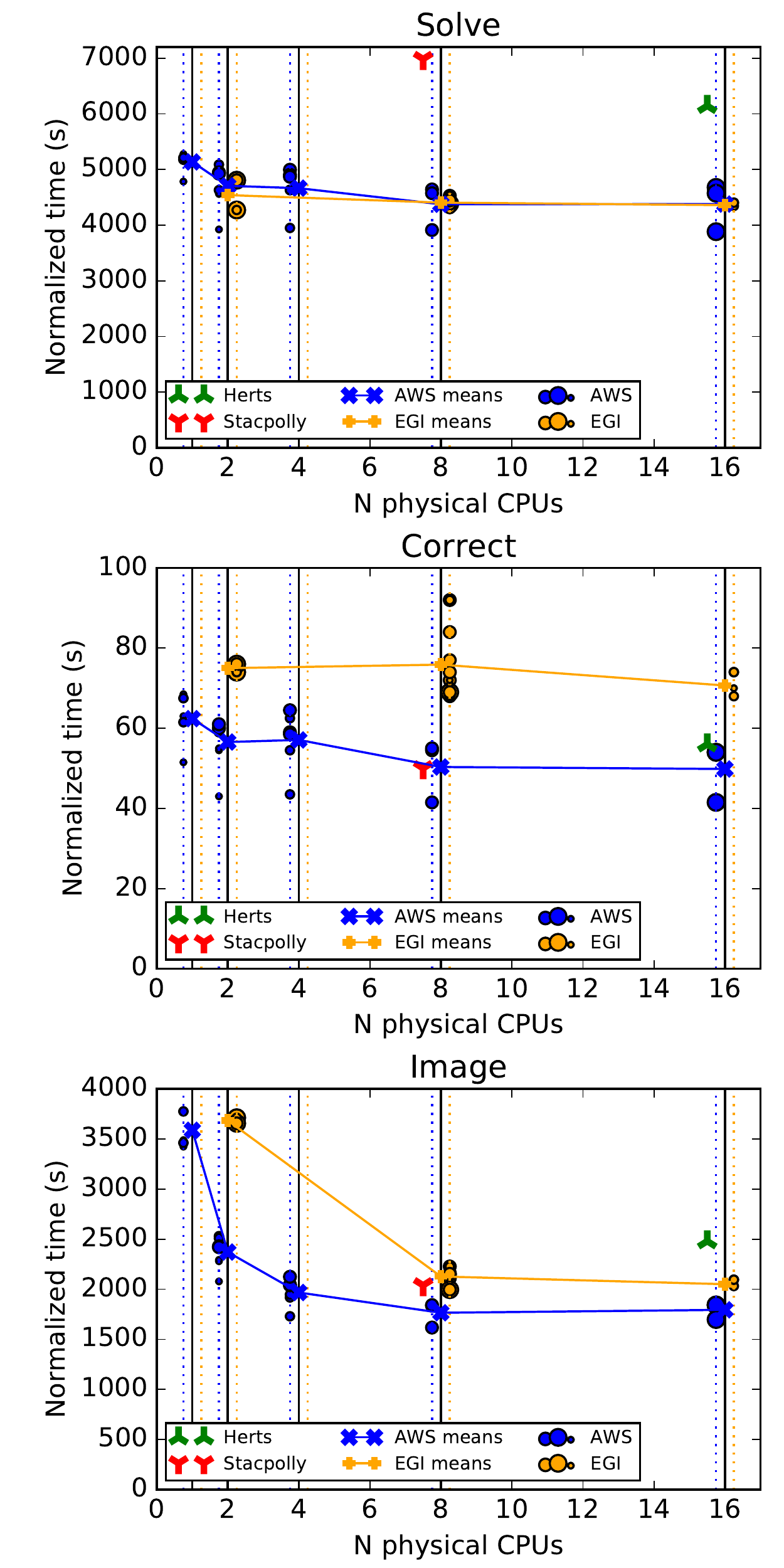}
 \caption{The time taken by the Solve, Correct and Image steps for our tests in 
different infrastructures. The CPU core numbers are normalized to physical CPUs 
(divided by a factor 2 in AWS). We show the individual tests in AWS (blue) and 
the EGI Federated Cloud (orange) cloud infrastructures as a cloud of points on 
either side of the main line that corresponds to the number of CPUs and the 
means in the centre. The size of the points is proportional to the memory of the 
instance used. As a comparison, we show the tests in the cluster Stacpolly in 
Edinburgh (red) and the LOFAR-UK Computing Facility in Hertfordshire (green). 
They are shown with a slight offset in x position for clarity.} 
\label{fig:times}
\end{figure}

\begin{figure*}
 \centering
 \includegraphics[width=\textwidth]{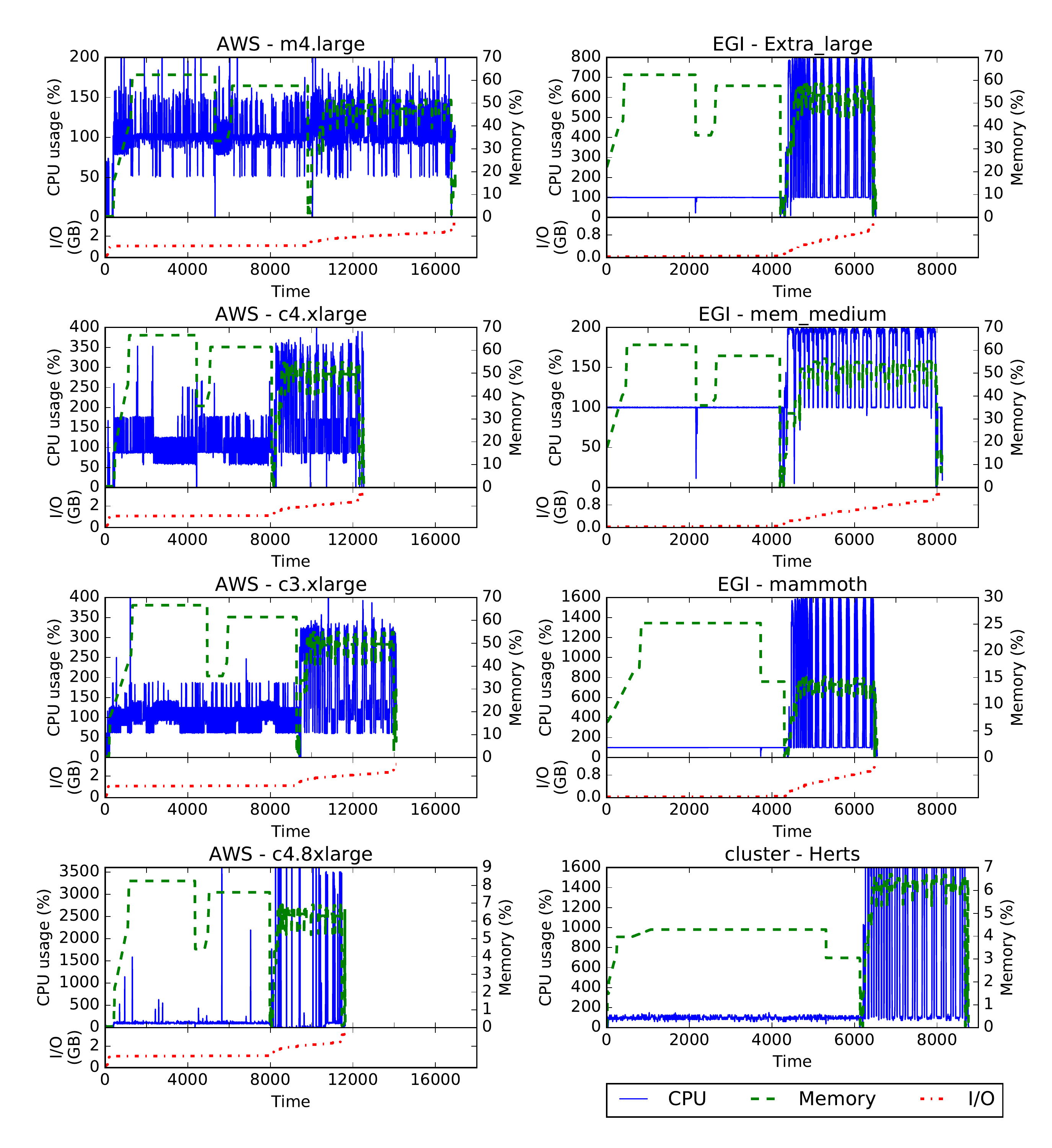}
 \caption{CPU and memory usage (blue solid and green dashed lines in the upper 
part of each panel) and input/output usage (red dash-dot line in lower part of 
each panel) for different instance types in the AWS and EGI Federated cloud 
infrastructures. The CPU and memory usage of the LOFAR-UK cluster in 
Hertfordshire is shown in the lower-right panel as a comparison. The AWS 
profiles include the initial data download step at the beginning (visible as a 
quick jump in the I/O usage). The ``solve'' and ``correct'' steps are at the 
beginning and consume approximately one core because they do not use 
multiprocessing. The ``solve'' step can be seen as the one or two flat peaks in 
memory usage and the ``correct'' is barely visible given it short duration. At 
the end, the ``image'' step is clearly visible with burst of parallel CPU usage 
and continuous I/O of data.}
 \label{fig:profile}
\end{figure*}

In Fig.~\ref{fig:times} we show the time taken in the ``solve'', ``correct'' 
and ``image'' steps of the pipeline. The number of CPU cores is normalized to 
physical cores and the time is also normalized. That means that the number of 
cores and times are divided by 2 for AWS data to allow a better comparison of 
the trends with the number of cores. We plot the data for the AWS and 
EGI Federated clouds and, as a comparison, the Stacpolly and LOFAR-UK
clusters. In Fig.~\ref{fig:profile} there are several panels showing the CPU 
and memory usage and the data I/O for different calibration test runs. In this 
figure it is possible to see the CPU and memory consumption of each step as 
well as the cumulative I/O.

In the ``solve'' step (upper panel in Fig.~\ref{fig:times}) the data is 
calibrated to an input model using BBS. The program uses only one core and it is 
not parallelized. This can be seen as the flat relation with the number of cores 
and the similar times in the two cloud infrastructures. The running times for 
the Stacpolly and Hertfordshire clusters are higher, at least in part due to the 
use of less powerful processors. This step could be treated as an embarrassingly 
parallel problem and several bands can be run in parallel in a single node or 
instance to maximize the performance. However, if the number of ``solve'' 
processes running in parallel in a single computing node is lower than the 
number of cores (for example, if there are less bands to process than cores) 
some of then will be idle with a corresponding waste of resources. In this case, 
the use of cloud instances with sizes adapted to the number of bands to process 
is one advantage of cloud infrastructures. The ``solve'' step can be seen in the 
initial part of the panels in Fig.~\ref{fig:profile} with one or two flat peaks 
of memory usage. In those cases the CPU core usage is approximately 100 per cent 
and the speed of the step is mainly determined by the CPU clock speed.

The ``correct'' step (middle panel in Fig.~\ref{fig:times}) involves applying 
the solutions computed in the ``solve'' step to the data and writing the 
corrected data back to the disk. It is computationally very simple and involves 
some data I/O. The times are very short in comparison with the ``solve'' step 
and the dispersion of the times is higher. In Fig.~\ref{fig:profile} the step is 
barely visible between the ``correct'' and ``image'' step given its short 
duration. The running times are very similar in the clusters and AWS and 
slightly higher in the EGI Federated Cloud, probably due to some overheads 
related to the use of an NFS file-system.

Finally, in the ``image'' step (lower panel in Fig.~\ref{fig:times}) the data is 
imaged using AWImager. Part of the imaging process is run in parallel making use 
of all the available processors. There is an improvement in the running time as 
the number of cores is increased. However, after a certain point there is no 
further significant improvement. Although the most computationally demanding 
part of the computing is parallelized, there is a part that runs in only one 
core. The time spent in this part cannot be improved using more cores and the 
time decreases asymptotically to the time spent on this part when we increase 
the number of cores \citep{AmdahlLaw}. This step can be seen in the last part of 
the panels shown in Fig.~\ref{fig:profile}. The parallel part uses all the CPU 
resources but the time spent in this part decreases as the number of cores 
increases while the single core part remains broadly constant.

\begin{figure}
 \centering
 \includegraphics[width=7.5cm]{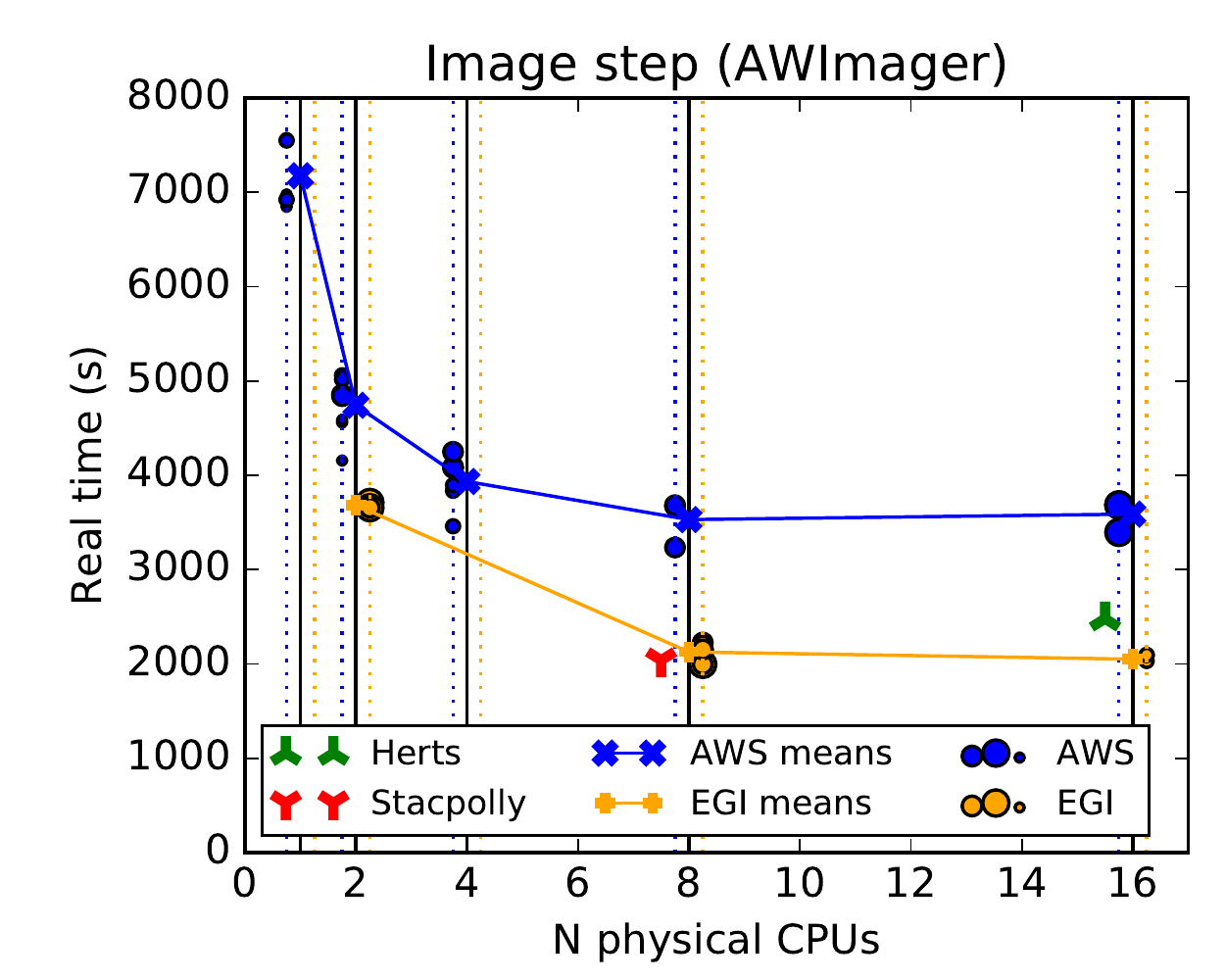}
 \caption{Time spent in the ``image'' step with respect to the number of 
physical cores of the instance or node. The symbols are as described in 
Fig.~\ref{fig:times}}
 \label{fig:realtimes}
\end{figure}

With the ``image'' step it is also possible to see the combined effect of 
hyperthreading and the lack of complete parallelization in a process. In 
Fig.~\ref{fig:realtimes} we show the real time spent in the step and the number 
of physical cores used. For the same number of physical cores we always get 
better performance in the EGI Federated Cloud infrastructure that does not use 
hyperthreading. The overheads found in AWS with respect to EGI are $29 \pm 8$ 
per cent with 2 physical cores; $66 \pm 13$ per cent with 8 physical cores; and, 
$75 \pm 9$ per cent with 16 physical cores. When hyperthreading is used, 
doubling the number of cores has little effect on the speed of the parallelized 
part (double the processors at half the speed) but the single core part runs at 
a lower speed (about half the speed). This can be seen in 
Fig.~\ref{fig:profile}, where the parallelized part (seen as peaks in the CPU 
usage in the last part of the profile) tends to be shorter with a higher number 
of cores while the single core part (the flat part with values close to 100 per 
cent between the peaks) remains similar in length. We observe that 
hyperthreading is only equivalent if several processes can be run in parallel 
until they use all the processor resources, or if the software runs fully in 
parallel. The difference in CPU usage paterns between the ``solve'', ``correct'' 
and ``image'' steps could be matched in a cloud infrastructure by a correct 
selection of the instance type. Additionally, if the cloud provides persistent 
block storage (like EBS in AWS), the single-core steps can be run in single-core 
instances and the parallelized steps in many-core instances working on the same 
data areas as required by the pipeline. This flexibility is one of the 
advantages of a cloud infrastructure.

\begin{figure}
 \centering
 \includegraphics[width=5.5cm]{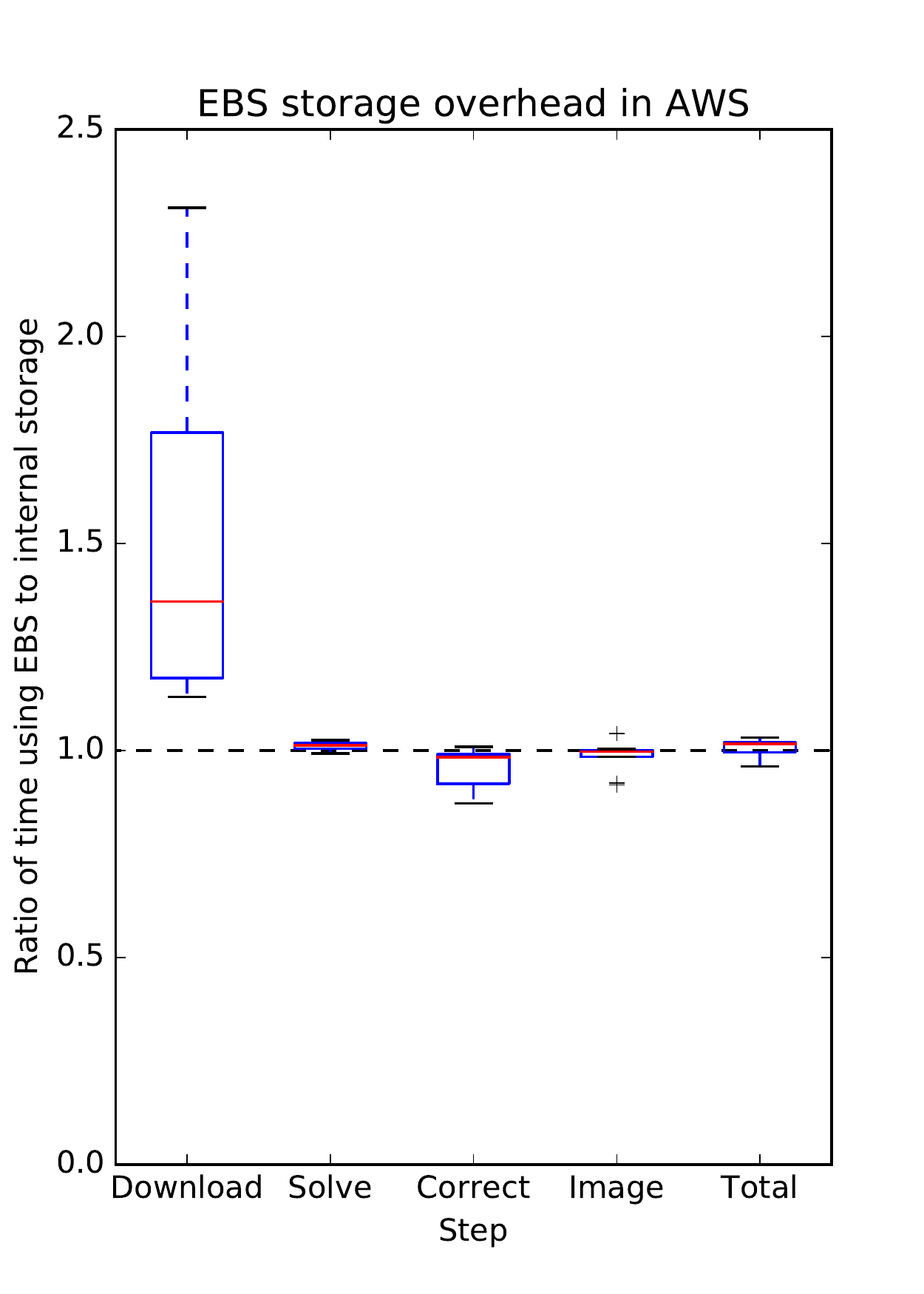}
 \caption{Relative time differences between EBS and internal storage volumes 
in the AWS infrastructure. The box and whiskers plots show the 
distribution of the fraction of time spent using an EBS volume 
with respect to using internal storage for each step.}
 \label{fig:io}
\end{figure}

For our tests, the effect of the memory available on the instance or node was 
not significant once a minimum threshold was reached. All the tests that run in 
instances with less than 8 GB of memory failed (bad\textunderscore alloc 
errors), but after this threshold was reached the performance was very similar, 
as shown in Fig.~\ref{fig:times}. The size of the scattered dots indicates the 
memory of the instance and, once the number of processors is fixed, only has a 
marginal effect on the time spent in the step. The memory threshold depends 
strongly on the size of the dataset. Tests with datasets that were 5 times 
bigger required a minimum of 32 GB of memory. The memory requirements depend 
strongly on the particular software used, the pipeline run, the number of 
processes running in parallel and the size of the dataset.

We also measured in AWS the performance of different steps (``download'', 
``solve'', ``correct'' and ``image'') in instances with different types of 
storage. By comparing the same instances using either internal storage or EBS 
storage we could determine if there is a significant overhead related to the use 
of EBS \citep{Dodson2016}. The results are shown in Fig.~\ref{fig:io}. The 
``download'' step is where the data were transferred from the object store to 
the internal storage and was only measured in AWS. Given the high bandwidth 
available, the time of the ``download'' step is mainly dominated by the I/O 
performance. In this case we found a median overhead of $\approx 36$ per cent 
over using internal storage. In the other steps there was no overhead within the 
error. We could not find a significant overhead in the overall test pipeline 
that depends on the use of EBS (the total overhead is $0.5 \pm 2.4 \%$).

One last point to consider in terms of performance is the effect of 
interruptions in the processing. With AWS spot instances the chance of a shut 
down of the instances increases with longer running times. Our simple tests were 
short but advanced pipelines take longer to run. In those cases we found it 
fundamental to be able to resume the running of the pipelines. Hence, it could 
be useful to design the pipeline software with well defined intermediate states 
from where it can resume the computing.

\subsection{Costs}
\label{costs}

We made an estimation of the costs of processing real LOFAR data in the AWS 
infrastructure versus using a dedicated cluster. There are many factors that can 
affect the accuracy of our estimate. One of the main factors is the optimization 
of the pipeline and infrastructure usage in AWS which is under development. For 
example, we are currently experimenting with improvements that can reduce the 
costs by more than 40 per cent, but we provide the current numbers as a general 
guideline. We do not consider in this comparison the price of long term object 
storage in the infrastructure (e.g. S3) and focus on the computing prices.

In a commercial cloud infrastructure the computing resources are consumed as 
they are used and they are usually charged at a unit rate per unit of time. In a 
dedicated cluster the initial price of the machine should be considered in 
addition to the running cost (electricity, support, etc). If the running cost of 
the dedicated cluster is lower than the cloud price, the time in which investing 
in a dedicated cluster pays off is given by the equation:
\begin{equation}
\label{eq:cloud_cluster}
 t\ p_{cloud} = c_{cluster} + t\ p_{cluster},
\end{equation} 
where $t$ is the time, $p_{cloud}$ is the cost of the cloud per unit of time, 
$c_{cluster}$ is the initial investment on the cluster, and  $p_{cluster}$ is 
the running cost of the cluster per unit of time.

For our estimation we used a real calibration run of a combination of 40 
sub-bands of one of the ELAIS-N1 datasets. The calibration followed the pipeline 
described in \citet{vanWeeren2016} and was run using the software called
\href{https://github.com/lofar-astron/factor}{\texttt{factor}}. It took 380 
hours to 
complete (just over two weeks). The final prices on AWS were \$\,177.75 for the 
computing with an \textit{m4.4xlarge} instance and \$\,157.77 for the EBS 
storage, adding up to a total of \$\,335.52, or \$\,0.883 per hour. The use of 
spot instances reduced the price from \$\,0.958 per hour to a mean of \$\,0.468 
per hour, that is 49 per cent of the nominal price of the instance. There was 
one interruption during the calibration process produced by the shut-down of the 
instance due to a spike in market prices. The \textit{m4.4xlarge} instance that 
we used in AWS has 16 vCPUs (8 physical cores of an Intel Xeon E5-2684 v4 at 
2.3 
GHz or E5-2676 v3 at 2.4 GHz), 64 GB of RAM memory and a 3 TB EBS solid state 
disk volume. We got a quote for a server with similar characteristics of 
$\approx \$\,9500$ (November 2016; exchange rate 1.25 USD to GBP). We considered 
in this case a running cost of \$\,4000 per year (\$\,1000 electricity charges 
and \$\,3000 for the human support). Using eq.~\ref{eq:cloud_cluster} we derived 
a time of 2.5 years for the costs of the two systems to be equivalent. 

We can estimate the number of datasets that can be processed during this time. 
In our case, the time expended was 380 hours for the 40 sub-bands, although the 
on-going efforts to optimise the software and selection of instances is 
continuing to improve this. The calibration of the 18 ELAIS-N1 full datasets 
would require to run 162 times this quantity of data. Hence, the computed 2.5 
years correspond to just over a third of the $162 \times 40$ sub-band runs.

The parameters used in our estimation can vary widely from case to case. For 
example, the cost of the support will be much lower if the computing resource is 
integrated into an existing cluster or if the electricity is already paid for 
the final user by third parties. We present in Table~\ref{table:prices} 
different possible scenarios and their estimated break-even times and number of 
full datasets that can be calibrated in this time obtained from 
eq.~\ref{eq:cloud_cluster}. 

We consider 4 different hardware infrastructures:  a) an 8-core node with 64 GB 
or memory and 3 TB of solid state disks for the scratch area with a network 
connection of 10 Gbit/s; b) the same option but with spinning disks; c) 16 cores 
with 128 GB of memory and 6 TB of solid state disks; and, d) 64 cores with 1 TB 
of memory and 12 TB of spinning disks. We also consider a range of operational 
scenarios which depend mainly on the context and support. The ``data centre'' 
scenario means that the user has access to a local data centre with specialised 
support in which to place the computing resource. This scenario provides a lower 
limit to the cost of the cluster with the electricity provided and a support 
cost of \$\,400 per year (based on a cost \$\,50 per year per core; E. Tittley 
private communication). In the ``no electricity'' scenario the support is 
estimated at \$\,3000 per year but the electricity is already provided by the 
centre. The ``base'' scenario is that considered above, where the user also has 
to pay the electricity consumed by the computing. In the ``alone'' scenario we 
raise the price of the support to \$\,10000 per year to account for the 
considerable amount of the time that the user or a third party must expend in 
maintaining the computing node or infrastructure. Finally, the ``dedicated'' 
scenario considers that one person (\$\,50000 per year) is fully dedicated to 
maintain the computing infrastructure. Although this scenario is not realistic, 
it marks an upper limit to the cost of the cluster.

\begin{table*}
\centering
\begin{minipage}{\textwidth}
\centering
\caption{Break-even times and the corresponding number of processed datasets  
estimated for different hardware infrastructures and operational scenarios.}
\label{table:prices}
\begin{tabular}{llllllllll}
\hline
\multicolumn{2}{c}{} & \multicolumn{8}{c}{Hardware infrastructure$^{1,2,3}$} \\
\cline{3-10}
\multicolumn{2}{c}{} & 
 \multicolumn{2}{l}{Normal (a)} & 
 \multicolumn{2}{l}{Cheap (b)} &  
 \multicolumn{2}{l}{$2\times$ (c)} & 
 \multicolumn{2}{l}{$8\times$ (d)}\\
\cline{3-10}
\multicolumn{2}{c}{} & 
 \multicolumn{2}{l}{$c_{cluster} = 9500$} & 
 \multicolumn{2}{l}{$c_{cluster} = 7000$} & 
 \multicolumn{2}{l}{$c_{cluster} = 17000$} & 
 \multicolumn{2}{l}{$c_{cluster} = 27000$} \\
\multicolumn{2}{c}{} & 
 \multicolumn{2}{l}{$p_{cloud} = 0.883$} & 
 \multicolumn{2}{l}{$p_{cloud} = 0.883$} & 
 \multicolumn{2}{l}{$p_{cloud} = 1.800$} & 
 \multicolumn{2}{l}{$p_{cloud} = 7.000$} \\
\multicolumn{2}{c}{Operational scenario$^{4,5}$} & 
 \multicolumn{2}{l}{$t_{band} = 380$} & 
 \multicolumn{2}{l}{$t_{band} = 600$} & 
 \multicolumn{2}{l}{$t_{band} = 170$} & 
 \multicolumn{2}{l}{$t_{band} = 60$} \\
\hline
 & $p_{cluster}$ & 
 t & n &  t & n &  t & n &  t & n \\
\hline
\hline
Data centre & 0.046 & 1.3 & 3.3 &  1.0 & 1.6 &  1.1 & 6.3 &  0.4 & 7.2\\
No electricity & 0.342 & 2.0 & 5.1 &  1.5 & 2.4 &  1.3 & 7.7 &  0.5 & 7.6\\
Base & 0.457 & 2.5 & 6.6 &  2.5 & 4.1 &  1.4 & 8.2 &  0.5 & 7.7\\
Alone & 1.256 & -- & -- &  -- & -- &  3.6 & 20.4 &  0.5 & 8.7\\
Dedicated & 5.822 & -- & -- &  -- & -- &  -- & -- &  2.6 & 42.4\\
\hline
\end{tabular}
\begin{list}{}{}
 \item $^1$ A detailed description of the hardware infrastructure can be 
found in the main text.
 \item $^2$ The units of $c_{cluster}$ are \$; for $p_{cloud}$ are \$ per hour; 
and, for $t_{band}$ are hours. A full dataset is composed of 9 bands.
 \item $^3$ The columns are: (1) t, time in years to break-even the costs 
of AWS and the cluster; (2) n, number of full datasets that can be calibrated 
in this time. A ``--'' indicates that the costs of AWS are always lower
for reasonable lifetimes (ie. t $>$ 5 yrs).
 \item $^4$ A detailed description of the operational scenarios can be found 
in the main text.
 \item $^5$ The units of $p_{cluster}$ are \$ per hour.
\end{list}
\end{minipage}
\end{table*}

As we can see in Table~\ref{table:prices} a bigger (dedicated) computing 
resource could outperform the AWS cloud in terms of costs. However, the size of 
the problem to solve must be relatively big (a considerable number of datasets 
need to be calibrated) and the money for such a resource must be available. For 
smaller and cheaper computing resources, the AWS cloud could offer a cheaper 
solution when the cost of the support is relatively high or the number of 
datasets to calibrate is not very high.

There are, however, several further differences to consider between the cluster 
and the cloud approach. A cloud infrastructure like AWS allows the parallel 
processing of several datasets, which may be interesting with tight deadlines. 
The obsolescence of the hardware was not taken into account in the previous 
calculations. The cost of AWS tends to go down steadily with time while the 
performance of the hardware is improved simultaneously. Furthermore, these 
calculations assume the cluster is used at 100 per cent efficiency: for 
intermittent use there will be wasted downtime, whilst on a commercial cloud 
you pay nothing for this.

 
\section{Review of infrastructure suitability}
\label{suitability}

In this section we combine our experiences and test results from our
investigations of different infrastructures, to present some general
considerations that are relevant for the calibration and analysis of LOFAR
(or other comparable) data in any computing platform. We then review the
issues specifically relevant to clouds.

\subsection{General considerations for all infrastructures}
\label{sect:general_considerations}

One of the main problems that can be found in any infrastructure occurs if there 
is a lack of support or documentation, or complexities in their usage. These 
factors have a great impact in the time consumed developing the calibration 
pipelines, to the extent of rendering some computing platforms practically 
unusable, particularly for a final user with limited time. The processes that 
were complex or unclear were especially time consuming requiring a lot of trial 
and error. This problem can be accentuated in bespoke solutions where the 
support of a wide community of users is not available. Good support, good and up 
to date documentation and the simplification of usage\footnote{``Simplicity is a 
great virtue but it requires hard work to achieve it and education to appreciate 
it. And to make matters worse: complexity sells better.'' \citep{EWD896}}, had a 
very positive impact on the developing time. 

The requirement of manual intervention for some tasks had a strong impact on the 
time spent. The difficulty of synchronizing agendas, the overload of support 
staff, the temporal unavailability of key contacts,  can all have an added 
negative impact when human intervention is required. This effect can add up 
quickly if many of the steps involved require this intervention or if it is 
required by simple tasks. This problem is minimized with the automation of the 
infrastructures. The time spent in this regard in fully automated 
infrastructures, like AWS, was minimal.

Regarding infrastructures, some of the main concerns in terms of performance are 
related to: the efficiency and configuration of the processing power; the memory 
available; the speed of the data transfer; and, the efficiency of the data I/O. 

In terms of processing speed, we found some effects that depended on the 
processor (processor speed, generation, etc.) but that were of second order. 
Multiprocessing was used in (parts of) tasks that were parallelized with the 
corresponding reduction of the processing time. The computational overhead due 
to virtualization in clouds was minimal (if the number of virtual CPU cores 
corresponded to the number of physical cores; we discuss the use of multiple 
virtual cores per physical core later). In general we found the cloud 
infrastructures tested to perform similarly or better than the clusters in terms 
of CPU speed. Many other factors had a much greater impact than this.

The memory was not a factor that limited processing performance, in general. The 
pipelines required a minimum amount of memory to run without failing. Once the 
instance or computing node provided this minimum amount, the availability of 
additional memory had little impact on the processing time of our test 
pipelines. The memory problems can be mitigated with new memory optimized 
software and a correct sizing of the resources.

The speed of the data transfer was initially thought to be a limiting factor but 
we did not find it to be a bottleneck on the overall process. The transfer times 
were lower than the processing times and orders of magnitude lower than the time 
spent developing the solutions in this early stage. During the data transfer 
step, the main overheads came from the requirement to manually supervise the 
transfer process (staging of data, creation and renewal of the proxy 
certificates, error checking, etc.) but not from the actual transfer of data.

The efficiency of the data I/O can have a potential impact on the speed of the 
calibration depending on the efficiency of the underlying system. This is 
particularly the case for our tests and some of the pipelines under development 
that are I/O bound instead of CPU bound. We did not find a large data I/O 
overhead coming from the use of non-local storage in the clouds. Specially, the 
use of EBS volumes in AWS had only a limited impact, less than $\approx 36$ per 
cent in the download of data step and practically null in the overall 
calibration tests. However, we found the lack of proper size scratch storage 
areas to be a limiting factor in many of the infrastructures that we tested. 

The availability and size of scratch storage areas turned out to be of great 
importance for the implementation of the pipelines. In many cases, we found that 
the pipelines required more scratch storage than the infrastructures were able 
to provide and this blocked the implementation. The Elastic Block Storage 
service of AWS proved to be very useful in this sense.

\subsection{Advantages and disadvantages of cloud infrastructures}
\label{sect:adv_cloud}

The cloud infrastructures excelled in the simplicity for the software 
installation and maintenance. Once the base operating system was ready, the 
installation of software and creation of instances with an exact copy of the 
software was straightforward. 

The availability of standard APIs and tools facilitated greatly the integration 
of the pipeline. Those APIs and tools (e.g. Ansible, Boto, or Ipython 
\citep[][]{ipython}) are widely used in the industry by communities out of the 
scientific circles which allows a transverse transfer of knowledge. The APIs of 
AWS or Open Stack were well documented and supported which permitted a quick 
integration of the software developed.

The flexibility of the cloud infrastructures allowed to adapt the hardware to 
the requirements of the pipelines, which is fundamental in the early stage of 
testing. Apart from that, this flexibility allows to optimize the consumption of 
resources once the calibration pipeline is well defined. All the cloud 
infrastructures permitted the parallel execution of the pipelines with a number 
of parallel instances that was only limited by the size of the cloud 
infrastructure or the resources allocated to the user. In the case of AWS the 
limits were very high allowing the parallel development of complex pipelines and 
the experimentation with different infrastructure parameters (instance size, 
volume sizes and types, etc.).

We found an issue that can be present in any infrastructure that uses 
virtualization but particularly on the cloud. The combination of Hyperthreading 
(where the number of processors is duplicated but the apparent clock speed is 
halved) with software that is not fully parallelized had a negative impact on 
the processing speed in the pipelines tested. The apparent lower speed per 
processor combined with the fact that parts of some processes did not, or could 
not, use multi-processing, caused an overhead in the overall run time \citep[see 
][]{AmdahlLaw}. Whilst hyperthreading may present advantages for other users, we 
found in our tests in AWS that the overhead in time was as high as 75 per cent 
in some steps of the pipeline. The performance would be lower overall but the 
real extent of the impact should be assessed with additional tests.

The model offered by the clouds, where the resources are consumed on-demand, 
could be of advantage in shared infrastructures where resources are allocated 
and consumed by a wide range of scientific projects and fields that may have 
different technical requirements. For a final user it offers the advantage of 
just paying for the usage of resources without the commitment required by 
dedicated infrastructures.

The costs associated to a commercial cloud like AWS can be equivalent or lower 
than those of a dedicated cluster in some common cases. It is a cost effective 
way of processing data if the user needs to calibrate low to medium volumes of 
data, or if they need otherwise to use a shared system in which the admin costs 
of maintaining up-to-date LOFAR software become high.

 
\section{Summary and conclusions}
\label{conclusions}

New scientific instruments are starting to generate an unprecedented amount of 
data. The capture, curation and analysis of these huge data volumes require the 
use of innovative strategies. The forthcoming Square Kilometre Array (SKA) will 
achieve data rates on a exabyte scale. LOFAR, one of the SKA pathfinders, is 
already producing data in a petabyte scale whose calibration present a 
formidable challenge. The several TB of data for each observation, a software 
whose installation and maintenance was not trivial, and a calibration pipeline 
that was quickly evolving and required intensive storage and computing 
resources, motivated us to investigate the use of different computing resources. 
After considering dedicated clusters and Grid infrastructures for the 
calibration of our LOFAR imaging data from the point of view of a final user, we 
focused mainly on cloud infrastructures. A cloud infrastructure can provide the 
flexibility and high throughput for the calibration of the big volumes of 
radio-astronomy data that we are handling.

Our initial tests on dedicated clusters were mainly limited by the complexity of 
the software installation and maintenance in these systems. An additional 
problem was the increase of computational resources required by the pipelines 
which rendered some of the clusters unable to run them. We also explored the use 
of Grid infrastructures but the need of manual intervention combined with the 
quick development of the pipelines at this early stage made them unsuitable for 
our needs. However, they are an infrastructure that should be evaluated once the 
pipelines are in a stable state. After that, we performed our tests in different 
cloud infrastructures: the EGI Federated Cloud, the RAL Cloud, and the 
commercial cloud Amazon Web Services. 

In general we found that good support, documentation, and simplicity of usage 
were of great importance for the implementation of the pipelines: the 
requirement of manual intervention had a strong negative impact on the time 
spent in some infrastructures. On the technical side, processing speed was 
comparable in the different infrastructures for similar resources. The quantity 
of memory had little impact on the processing speed once a minimum amount of 
required memory was available. The speed of data transfer was not one of the 
main limiting factors as the transfer time was lower than the computing time. 
Finally, we could not find a strong data I/O overhead coming from the use of 
non-local storage in the clouds.

However, we identified a couple of unforeseen issues that had a negative impact 
on the implementation of pipelines. The combination of hyperthreading and tasks 
of the pipelines that are not, or cannot be, fully parallelized produced an 
empirical overhead in the running time of the pipelines.  Additionally, the lack 
of scratch storage areas of an appropriate size could block the implementation 
of the pipelines in some systems.

Cloud infrastructures presented several highlights, most notably: a) the 
straightforward and simple installation and maintenance of the software; b) the 
availability of standard APIs and tools widely used in the industry; c) the 
flexibility to adapt the infrastructure to the needs of the problem; and, d) the 
on-demand consumption of shared resources.

We found that the run of data calibration pipelines is not just possible but 
efficient in cloud infrastructures. From the point of view of the final user it 
simplified many important steps and solved issues that blocked the 
implementation or running of the pipelines in other infrastructures. In terms of 
costs, a commercial cloud infrastructure like AWS is currently worthwhile in 
several common use cases, where the user lacks access to powerful storage and 
computing resources or specialised support, or where the calibration of small to 
medium sets of data is needed. 

In the future we will present the detailed results and technical details of the 
calibration of the LOFAR ELAIS-N1 data on AWS. We are currently 
optimising and integrating in AWS the full calibration pipeline 
presented in \citet{vanWeeren2016}\footnote{The current efforts can be found 
in \href{http://www.lofarcloud.uk}{http://www.lofarcloud.uk}}. We also plan to 
perform tests in other clouds and continue with the study of the integration in 
the EGI and RAL clouds. We will test the use of COMPs and, once the pipeline is 
stable, the integration of the full pipeline in a Grid infrastructure will be 
considered.

\section*{Acknowledgements}
We acknowledge the useful comments of the anonymous referee.
We would like to acknowledge the work of all the developers and packagers 
of the LOFAR software that constitute the core of the 
processing pipelines (including 
\href{https://github.com/lofar-astron/factor}{factor}, 
\href{https://github.com/lofar-astron/prefactor}{pre-factor}, 
\href{https://github.com/darafferty/LSMTool}{LSMTool}, 
\href{https://github.com/revoltek/losoto}{LoSoTo}, and the 
\href{http://kernsuite.info/}{Kern Suite}), as well as the useful discussions 
with the participants 
in the LOFAR blank fields and direction dependent calibration teleconferences 
over the years.
JS and PNB are grateful for financial support from STFC via grant ST/M001229/1.
This work has been also supported by the projects `AMIGA5: gas in and around
galaxies. Scientific and technological preparation for the SKA'
(AYA2014-52013-C2-1-R) and `AMIGA6: gas in and around
galaxies. Preparation for SKA science and contribution to the design of
the SKA data flow' (AYA2015-65973-C3-1-R) both of which were
co-funded by MICINN and FEDER funds and the Junta de Andaluc\'{\i}a
(Spain) grants P08-FQM-4205 and TIC-114.
We would like to explicitly acknowledge Dr Jose 
Ruedas -- chief of the computer centre and responsible of the computing and 
communications infrastructures at IAA-CSIC -- and Rafael Parra -- system 
administrator of the IAA computing cluster -- for their technical 
assistance.
We acknowledge the joint SKA and AWS Astrocompute proposal call that 
was used to fund all the tests in the AWS infrastructure with the projects 
``Calibration of LOFAR ELAIS-N1 data in the Amazon cloud'' and ``Amazon Cloud 
Processing of LOFAR Tier-1 surveys: Opening up a new window on the Universe''.
This work made use of the University of Hertfordshire's 
\href{http://stri-cluster.herts.ac.uk/}{high-performance computing facility} 
and the LOFAR-UK computing facility, supported by STFC [grant number
ST/P000096/1].
This work benefited from services and resources provided by the fedcloud.egi.eu 
Virtual Organization, supported by the national resource providers of the EGI 
Federation.
We acknowledge the resources and support provided by the STFC RAL Cloud 
infrastructure.
LOFAR, the Low Frequency Array designed and constructed by ASTRON, has 
facilities in several countries, that are owned by various parties (each with 
their own funding sources), and that are collectively operated by the 
International LOFAR Telescope (ILT) foundation under a joint scientific policy.

\bibliographystyle{model2-names-jsm}
\bibliography{databasesol}

\begin{thebibliography}{32}
\expandafter\ifx\csname natexlab\endcsname\relax\def\natexlab#1{#1}\fi
\providecommand{\url}[1]{\texttt{#1}}
\providecommand{\href}[2]{#2}
\providecommand{\path}[1]{#1}
\providecommand{\DOIprefix}{doi:}
\providecommand{\ArXivprefix}{arXiv:}
\providecommand{\URLprefix}{URL: }
\providecommand{\Pubmedprefix}{pmid:}
\providecommand{\doi}[1]{\href{http://dx.doi.org/#1}{\path{#1}}}
\providecommand{\Pubmed}[1]{\href{pmid:#1}{\path{#1}}}
\providecommand{\bibinfo}[2]{#2}
\ifx\xfnm\relax \def\xfnm[#1]{\unskip,\space#1}\fi
\bibitem[{Altschul et~al.(1990)Altschul, Gish, Miller, Myers and
  Lipman}]{Altschul1990}
\bibinfo{author}{Altschul S.~F.}, \bibinfo{author}{Gish W.},
  \bibinfo{author}{Miller W.}, \bibinfo{author}{Myers E.~W.},
  \bibinfo{author}{Lipman D.~J.}, \bibinfo{year}{1990}.
\newblock \bibinfo{title}{Basic local alignment search tool}.
\newblock \bibinfo{journal}{Journal of Molecular Biology}
  \bibinfo{volume}{215}, \bibinfo{pages}{403 -- 410}.
\newblock \URLprefix
  \url{http://www.sciencedirect.com/science/article/pii/S0022283605803602},
  \DOIprefix\doi{http://dx.doi.org/10.1016/S0022-2836(05)80360-2}.
\bibitem[{Amdahl(2007)}]{AmdahlLaw}
\bibinfo{author}{Amdahl G.~M.}, \bibinfo{year}{2007}.
\newblock \bibinfo{title}{Validity of the single processor approach to
  achieving large scale computing capabilities, reprinted from the afips
  conference proceedings, vol. 30 (atlantic city, n.j., apr. 18–20), afips
  press, reston, va., 1967, pp. 483–485, when dr. amdahl was at international
  business machines corporation, sunnyvale, california}.
\newblock \bibinfo{journal}{IEEE Solid-State Circuits Society Newsletter}
  \bibinfo{volume}{12}, \bibinfo{pages}{19--20}.
\newblock \DOIprefix\doi{10.1109/N-SSC.2007.4785615}.
\bibitem[{Berriman et~al.(2012)Berriman, Deelman, Juve, Rynge and
  V{\"o}ckler}]{Berriman2012}
\bibinfo{author}{Berriman G.~B.}, \bibinfo{author}{Deelman E.},
  \bibinfo{author}{Juve G.}, \bibinfo{author}{Rynge M.},
  \bibinfo{author}{V{\"o}ckler J.~S.}, \bibinfo{year}{2012}.
\newblock \bibinfo{title}{The application of cloud computing to scientific
  workflows: a study of cost and performance}.
\newblock \bibinfo{journal}{Philosophical Transactions of the Royal Society of
  London A: Mathematical, Physical and Engineering Sciences}
  \bibinfo{volume}{371}.
\newblock \DOIprefix\doi{10.1098/rsta.2012.0066}.
\bibitem[{Boettiger(2015)}]{Boettiger2015}
\bibinfo{author}{Boettiger C.}, \bibinfo{year}{2015}.
\newblock \bibinfo{title}{An introduction to docker for reproducible research},
  pp. \bibinfo{pages}{71--79}.
\newblock \DOIprefix\doi{10.1145/2723872.2723882}. \bibinfo{note}{cited By 27}.
\bibitem[{{Breitling} et~al.(2015){Breitling}, {Mann}, {Vocks}, {Steinmetz} and
  {Strassmeier}}]{Breitling2015}
\bibinfo{author}{{Breitling} F.}, \bibinfo{author}{{Mann} G.},
  \bibinfo{author}{{Vocks} C.}, \bibinfo{author}{{Steinmetz} M.},
  \bibinfo{author}{{Strassmeier} K.~G.}, \bibinfo{year}{2015}.
\newblock \bibinfo{title}{{The LOFAR Solar Imaging Pipeline and the LOFAR Solar
  Data Center}}.
\newblock \bibinfo{journal}{Astronomy and Computing} \bibinfo{volume}{13},
  \bibinfo{pages}{99--107}.
\newblock \DOIprefix\doi{10.1016/j.ascom.2015.08.001},
  \href{http://arxiv.org/abs/1603.05990}{\tt arXiv:1603.05990}.
\bibitem[{Dijkstra(1984)}]{EWD896}
\bibinfo{author}{Dijkstra E.~W.}, \bibinfo{year}{1984}.
\newblock \bibinfo{title}{On the nature of computing science}.
\newblock \URLprefix
  \url{http://www.cs.utexas.edu/users/EWD/ewd08xx/EWD896.PDF}.
  \bibinfo{note}{circulated privately}.
\bibitem[{{Dodson} et~al.(2016){Dodson}, {Vinsen}, {Wu}, {Popping}, {Meyer},
  {Wicenec}, {Quinn}, {van Gorkom} and {Momjian}}]{Dodson2016}
\bibinfo{author}{{Dodson} R.} et~al., \bibinfo{year}{2016}.
\newblock \bibinfo{title}{{Imaging SKA-scale data in three different computing
  environments}}.
\newblock \bibinfo{journal}{Astronomy and Computing} \bibinfo{volume}{14},
  \bibinfo{pages}{8--22}.
\newblock \DOIprefix\doi{10.1016/j.ascom.2015.10.007},
  \href{http://arxiv.org/abs/1511.00401}{\tt arXiv:1511.00401}.
\bibitem[{{Ekers}(2012)}]{Ekers2012}
\bibinfo{author}{{Ekers} R.}, \bibinfo{year}{2012}.
\newblock \bibinfo{title}{{The History of the Square Kilometre Array (SKA) -
  Born Global}}.
\newblock \bibinfo{journal}{ArXiv e-prints}
  \href{http://arxiv.org/abs/1212.3497}{\tt arXiv:1212.3497}.
\bibitem[{{Hardcastle} et~al.(2016){Hardcastle}, {G{\"u}rkan}, {van Weeren},
  {Williams}, {Best}, {de Gasperin}, {Rafferty}, {Read}, {Sabater}, {Shimwell},
  {Smith}, {Tasse}, {Bourne}, {Brienza}, {Br{\"u}ggen}, {Brunetti},
  {Chy{\.z}y}, {Conway}, {Dunne}, {Eales}, {Maddox}, {Jarvis}, {Mahony},
  {Morganti}, {Prandoni}, {R{\"o}ttgering}, {Valiante} and
  {White}}]{Hardcastle2016}
\bibinfo{author}{{Hardcastle} M.~J.} et~al., \bibinfo{year}{2016}.
\newblock \bibinfo{title}{{LOFAR/H-ATLAS: a deep low-frequency survey of the
  Herschel-ATLAS North Galactic Pole field}}.
\newblock \bibinfo{journal}{\mnras} \bibinfo{volume}{462},
  \bibinfo{pages}{1910--1936}.
\newblock \DOIprefix\doi{10.1093/mnras/stw1763},
  \href{http://arxiv.org/abs/1606.09437}{\tt arXiv:1606.09437}.
\bibitem[{{Heald} et~al.(2015){Heald}, {Pizzo}, {Orr{\'u}}, {Breton},
  {Carbone}, {Ferrari}, {Hardcastle}, {Jurusik}, {Macario}, {Mulcahy},
  {Rafferty}, {Asgekar}, {Brentjens}, {Fallows}, {Frieswijk}, {Toribio},
  {Adebahr}, {Arts}, {Bell}, {Bonafede}, {Bray}, {Broderick}, {Cantwell},
  {Carroll}, {Cendes}, {Clarke}, {Croston}, {Daiboo}, {de Gasperin}, {Gregson},
  {Harwood}, {Hassall}, {Heesen}, {Horneffer}, {van der Horst}, {Iacobelli},
  {Jeli{\'c}}, {Jones}, {Kant}, {Kokotanekov}, {Martin}, {McKean}, {Morabito},
  {Nikiel-Wroczy{\'n}ski}, {Offringa}, {Pandey}, {Pandey-Pommier}, {Pietka},
  {Pratley}, {Riseley}, {Rowlinson}, {Sabater}, {Scaife}, {Scheers},
  {Sendlinger}, {Shulevski}, {Sipior}, {Sobey}, {Stewart}, {Stroe}, {Swinbank},
  {Tasse}, {Tr{\"u}stedt}, {Varenius}, {van Velzen}, {Vilchez}, {van Weeren},
  {Wijnholds}, {Williams}, {de Bruyn}, {Nijboer}, {Wise}, {Alexov}, {Anderson},
  {Avruch}, {Beck}, {Bell}, {van Bemmel}, {Bentum}, {Bernardi}, {Best},
  {Breitling}, {Brouw}, {Br{\"u}ggen}, {Butcher}, {Ciardi}, {Conway}, {de
  Geus}, {de Jong}, {de Vos}, {Deller}, {Dettmar}, {Duscha}, {Eisl{\"o}ffel},
  {Engels}, {Falcke}, {Fender}, {Garrett}, {Grie{\ss}meier}, {Gunst},
  {Hamaker}, {Hessels}, {Hoeft}, {H{\"o}randel}, {Holties}, {Intema},
  {Jackson}, {J{\"u}tte}, {Karastergiou}, {Klijn}, {Kondratiev}, {Koopmans},
  {Kuniyoshi}, {Kuper}, {Law}, {van Leeuwen}, {Loose}, {Maat}, {Markoff},
  {McFadden}, {McKay-Bukowski}, {Mevius}, {Miller-Jones}, {Morganti}, {Munk},
  {Nelles}, {Noordam}, {Norden}, {Paas}, {Polatidis}, {Reich}, {Renting},
  {R{\"o}ttgering}, {Schoenmakers}, {Schwarz}, {Sluman}, {Smirnov}, {Stappers},
  {Steinmetz}, {Tagger}, {Tang}, {ter Veen}, {Thoudam}, {Vermeulen}, {Vocks},
  {Vogt}, {Wijers}, {Wucknitz}, {Yatawatta} and {Zarka}}]{Heald2015}
\bibinfo{author}{{Heald} G.~H.} et~al., \bibinfo{year}{2015}.
\newblock \bibinfo{title}{{The LOFAR Multifrequency Snapshot Sky Survey (MSSS).
  I. Survey description and first results}}.
\newblock \bibinfo{journal}{\aap} \bibinfo{volume}{582}, \bibinfo{pages}{A123}.
\newblock \DOIprefix\doi{10.1051/0004-6361/201425210},
  \href{http://arxiv.org/abs/1509.01257}{\tt arXiv:1509.01257}.
\bibitem[{{Holties} et~al.(2012){Holties}, {Renting} and
  {Grange}}]{Holties2012}
\bibinfo{author}{{Holties} H.}, \bibinfo{author}{{Renting} A.},
  \bibinfo{author}{{Grange} Y.}, \bibinfo{year}{2012}.
\newblock \bibinfo{title}{{The LOFAR long-term archive: e-infrastructure on
  petabyte scale}}, in: \bibinfo{booktitle}{Software and Cyberinfrastructure
  for Astronomy II}, p. \bibinfo{pages}{845117}.
\newblock \DOIprefix\doi{10.1117/12.927147}.
\bibitem[{{Intema} et~al.(2009){Intema}, {van der Tol}, {Cotton}, {Cohen}, {van
  Bemmel} and {R{\"o}ttgering}}]{Intema2009}
\bibinfo{author}{{Intema} H.~T.}, \bibinfo{author}{{van der Tol} S.},
  \bibinfo{author}{{Cotton} W.~D.}, \bibinfo{author}{{Cohen} A.~S.},
  \bibinfo{author}{{van Bemmel} I.~M.}, \bibinfo{author}{{R{\"o}ttgering}
  H.~J.~A.}, \bibinfo{year}{2009}.
\newblock \bibinfo{title}{{Ionospheric calibration of low frequency radio
  interferometric observations using the peeling scheme. I. Method description
  and first results}}.
\newblock \bibinfo{journal}{\aap} \bibinfo{volume}{501},
  \bibinfo{pages}{1185--1205}.
\newblock \DOIprefix\doi{10.1051/0004-6361/200811094},
  \href{http://arxiv.org/abs/0904.3975}{\tt arXiv:0904.3975}.
\bibitem[{ISO/IEC 17826:2016()}]{CDMI}
ISO/IEC 17826:2016, \bibinfo{year}{2016}.
\newblock \bibinfo{title}{{Information technology -- Cloud Data Management
  Interface (CDMI)}}.
\bibitem[{{Kemball} and {Wieringa}(2000)}]{MS_definition}
\bibinfo{author}{{Kemball} A.~J.}, \bibinfo{author}{{Wieringa} M.~H.},
  \bibinfo{year}{2000}.
\newblock \bibinfo{title}{Measurementset definition version 2.0}.
\newblock \URLprefix \url{http://casa.nrao.edu/Memos/229.html}.
\bibitem[{L\"{a}mmel(2008)}]{Lammel2008}
\bibinfo{author}{L\"{a}mmel R.}, \bibinfo{year}{2008}.
\newblock \bibinfo{title}{Google’s mapreduce programming model —
  revisited}.
\newblock \bibinfo{journal}{Science of Computer Programming}
  \bibinfo{volume}{70}, \bibinfo{pages}{1 -- 30}.
\newblock \URLprefix
  \url{http://www.sciencedirect.com/science/article/pii/S0167642307001281},
  \DOIprefix\doi{http://dx.doi.org/10.1016/j.scico.2007.07.001}.
\bibitem[{Lordan et~al.(2014)Lordan, Tejedor, Ejarque, Rafanell, {\'A}lvarez,
  Marozzo, Lezzi, Sirvent, Talia and Badia}]{Lordan2014}
\bibinfo{author}{Lordan F.} et~al., \bibinfo{year}{2014}.
\newblock \bibinfo{title}{Servicess: An interoperable programming framework for
  the cloud}.
\newblock \bibinfo{journal}{Journal of Grid Computing} \bibinfo{volume}{12},
  \bibinfo{pages}{67--91}.
\newblock \URLprefix \url{http://dx.doi.org/10.1007/s10723-013-9272-5},
  \DOIprefix\doi{10.1007/s10723-013-9272-5}.
\bibitem[{Marr et~al.(2002)Marr, Binns, Hill, Hinton, Koufaty, Miller and
  Upton}]{Marr2002}
\bibinfo{author}{Marr D.~T.}, \bibinfo{author}{Binns F.}, \bibinfo{author}{Hill
  D.~L.}, \bibinfo{author}{Hinton G.}, \bibinfo{author}{Koufaty D.~A.},
  \bibinfo{author}{Miller J.~A.}, \bibinfo{author}{Upton M.},
  \bibinfo{year}{2002}.
\newblock \bibinfo{title}{Hyper-threading technology architecture and
  microarchitecture}.
\newblock \bibinfo{journal}{j-INTEL-TECH-J} \bibinfo{volume}{6},
  \bibinfo{pages}{4--15}.
\newblock \URLprefix
  \url{http://developer.intel.com/technology/itj/2002/volume06issue01/vol6iss1_hyper_threading_technology.pdf}.
\bibitem[{{McMullin} et~al.(2007){McMullin}, {Waters}, {Schiebel}, {Young} and
  {Golap}}]{McMullin2007}
\bibinfo{author}{{McMullin} J.~P.}, \bibinfo{author}{{Waters} B.},
  \bibinfo{author}{{Schiebel} D.}, \bibinfo{author}{{Young} W.},
  \bibinfo{author}{{Golap} K.}, \bibinfo{year}{2007}.
\newblock \bibinfo{title}{{CASA Architecture and Applications}}, in:
  \bibinfo{editor}{{Shaw} R.~A.}, \bibinfo{editor}{{Hill} F.},
  \bibinfo{editor}{{Bell} D.~J.} (Eds.), \bibinfo{booktitle}{Astronomical Data
  Analysis Software and Systems XVI}, p. \bibinfo{pages}{127}.
\bibitem[{{Norris} et~al.(2013){Norris}, {Afonso}, {Bacon}, {Beck}, {Bell},
  {Beswick}, {Best}, {Bhatnagar}, {Bonafede}, {Brunetti}, {Budav{\'a}ri},
  {Cassano}, {Condon}, {Cress}, {Dabbech}, {Feain}, {Fender}, {Ferrari},
  {Gaensler}, {Giovannini}, {Haverkorn}, {Heald}, {Van der Heyden}, {Hopkins},
  {Jarvis}, {Johnston-Hollitt}, {Kothes}, {Van Langevelde}, {Lazio}, {Mao},
  {Mart{\'{\i}}nez-Sansigre}, {Mary}, {Mcalpine}, {Middelberg}, {Murphy},
  {Padovani}, {Paragi}, {Prandoni}, {Raccanelli}, {Rigby}, {Roseboom},
  {R{\"o}ttgering}, {Sabater}, {Salvato}, {Scaife}, {Schilizzi}, {Seymour},
  {Smith}, {Umana}, {Zhao} and {Zinn}}]{Norris2013}
\bibinfo{author}{{Norris} R.~P.} et~al., \bibinfo{year}{2013}.
\newblock \bibinfo{title}{{Radio Continuum Surveys with Square Kilometre Array
  Pathfinders}}.
\newblock \bibinfo{journal}{\pasa} \bibinfo{volume}{30}, \bibinfo{pages}{20}.
\newblock \DOIprefix\doi{10.1017/pas.2012.020},
  \href{http://arxiv.org/abs/1210.7521}{\tt arXiv:1210.7521}.
\bibitem[{{Pandey} et~al.(2009){Pandey}, {van Zwieten}, {de Bruyn} and
  {Nijboer}}]{BBS}
\bibinfo{author}{{Pandey} V.~N.}, \bibinfo{author}{{van Zwieten} J.~E.},
  \bibinfo{author}{{de Bruyn} A.~G.}, \bibinfo{author}{{Nijboer} R.},
  \bibinfo{year}{2009}.
\newblock \bibinfo{title}{{Calibrating LOFAR using the Black Board Selfcal
  System}}, in: \bibinfo{editor}{{Saikia} D.~J.}, \bibinfo{editor}{{Green}
  D.~A.}, \bibinfo{editor}{{Gupta} Y.}, \bibinfo{editor}{{Venturi} T.} (Eds.),
  \bibinfo{booktitle}{The Low-Frequency Radio Universe}, p.
  \bibinfo{pages}{384}.
\bibitem[{P\'erez and Granger(2007)}]{ipython}
\bibinfo{author}{P\'erez F.}, \bibinfo{author}{Granger B.~E.},
  \bibinfo{year}{2007}.
\newblock \bibinfo{title}{{IP}ython: a system for interactive scientific
  computing}.
\newblock \bibinfo{journal}{Computing in Science and Engineering}
  \bibinfo{volume}{9}, \bibinfo{pages}{21--29}.
\newblock \URLprefix \url{http://ipython.org},
  \DOIprefix\doi{10.1109/MCSE.2007.53}.
\bibitem[{{R{\"o}ttgering} et~al.(2011){R{\"o}ttgering}, {Afonso}, {Barthel},
  {Batejat}, {Best}, {Bonafede}, {Br{\"u}ggen}, {Brunetti}, {Chy{\.z}y},
  {Conway}, {Gasperin}, {Ferrari}, {Haverkorn}, {Heald}, {Hoeft}, {Jackson},
  {Jarvis}, {Ker}, {Lehnert}, {Macario}, {McKean}, {Miley}, {Morganti},
  {Oosterloo}, {Orr{\`u}}, {Pizzo}, {Rafferty}, {Shulevski}, {Tasse}, {Bemmel},
  {van der Tol}, {van Weeren}, {Verheijen}, {White} and
  {Wise}}]{Rottgering2011}
\bibinfo{author}{{R{\"o}ttgering} H.} et~al., \bibinfo{year}{2011}.
\newblock \bibinfo{title}{{LOFAR and APERTIF Surveys of the Radio Sky: Probing
  Shocks and Magnetic Fields in Galaxy Clusters}}.
\newblock \bibinfo{journal}{Journal of Astrophysics and Astronomy}
  \bibinfo{volume}{32}, \bibinfo{pages}{557--566}.
\newblock \DOIprefix\doi{10.1007/s12036-011-9129-x},
  \href{http://arxiv.org/abs/1107.1606}{\tt arXiv:1107.1606}.
\bibitem[{Sandberg(1986)}]{Sandberg1986}
\bibinfo{author}{Sandberg R.}, \bibinfo{year}{1986}.
\newblock \bibinfo{title}{The Sun Network File System: Design, Implementation
  and Experience}.
\newblock \bibinfo{type}{Technical Report}. in Proceedings of the Summer 1986
  USENIX Technical Conference and Exhibition.
\bibitem[{{Shimwell} et~al.(2016a){Shimwell}, {Luckin}, {Br{\"u}ggen},
  {Brunetti}, {Intema}, {Owers}, {R{\"o}ttgering}, {Stroe}, {van Weeren},
  {Williams}, {Cassano}, {de Gasperin}, {Heald}, {Hoang}, {Hardcastle},
  {Sridhar}, {Sabater}, {Best}, {Bonafede}, {Chy{\.z}y}, {En{\ss}lin},
  {Ferrari}, {Haverkorn}, {Hoeft}, {Horellou}, {McKean}, {Morabito},
  {Orr{\`u}}, {Pizzo}, {Retana-Montenegro} and {White}}]{Shimwell2016}
\bibinfo{author}{{Shimwell} T.~W.} et~al., \bibinfo{year}{2016}a.
\newblock \bibinfo{title}{{A plethora of diffuse steep spectrum radio sources
  in Abell 2034 revealed by LOFAR}}.
\newblock \bibinfo{journal}{\mnras} \bibinfo{volume}{459},
  \bibinfo{pages}{277--290}.
\newblock \DOIprefix\doi{10.1093/mnras/stw661},
  \href{http://arxiv.org/abs/1603.06591}{\tt arXiv:1603.06591}.
\bibitem[{{Shimwell} et~al.(2016b){Shimwell}, {R{\"o}ttgering}, {Best},
  {Williams}, {Dijkema}, {de Gasperin}, {Hardcastle}, {Heald}, {Hoang},
  {Horneffer}, {Intema}, {Mahony}, {Mandal}, {Mechev}, {Morabito}, {Oonk},
  {Rafferty}, {Retana-Montenegro}, {Sabater}, {Tasse}, {van Weeren},
  {Br{\"u}ggen}, {Brunetti}, {Chy{\.z}y}, {Conway}, {Haverkorn}, {Jackson},
  {Jarvis}, {McKean}, {Miley}, {Morganti}, {White}, {Wise}, {van Bemmel},
  {Beck}, {Brienza}, {Bonafede}, {Calistro Rivera}, {Cassano}, {Clarke},
  {Cseh}, {Deller}, {Drabent}, {van Driel}, {Engels}, {Falcke}, {Ferrari},
  {Fr{\"o}hlich}, {Garrett}, {Harwood}, {Heesen}, {Hoeft}, {Horellou},
  {Israel}, {Kapi{\'n}ska}, {Kunert-Bajraszewska}, {McKay}, {Mohan},
  {Orr{\'u}}, {Pizzo}, {Prandoni}, {Schwarz}, {Shulevski}, {Sipior}, {Smith},
  {Sridhar}, {Steinmetz}, {Stroe}, {Varenius}, {van der Werf}, {Zensus} and
  {Zwart}}]{Shimwell2016barXiv}
\bibinfo{author}{{Shimwell} T.~W.} et~al., \bibinfo{year}{2016}b.
\newblock \bibinfo{title}{{The LOFAR Two-metre Sky Survey - I. Survey
  Description and Preliminary Data Release}}.
\newblock \bibinfo{journal}{ArXiv e-prints}
  \href{http://arxiv.org/abs/1611.02700}{\tt arXiv:1611.02700}.
\bibitem[{{Tasse} et~al.(2013){Tasse}, {van der Tol}, {van Zwieten}, {van
  Diepen} and {Bhatnagar}}]{Tasse2013}
\bibinfo{author}{{Tasse} C.}, \bibinfo{author}{{van der Tol} S.},
  \bibinfo{author}{{van Zwieten} J.}, \bibinfo{author}{{van Diepen} G.},
  \bibinfo{author}{{Bhatnagar} S.}, \bibinfo{year}{2013}.
\newblock \bibinfo{title}{{Applying full polarization A-Projection to very wide
  field of view instruments: An imager for LOFAR}}.
\newblock \bibinfo{journal}{\aap} \bibinfo{volume}{553}, \bibinfo{pages}{A105}.
\newblock \DOIprefix\doi{10.1051/0004-6361/201220882},
  \href{http://arxiv.org/abs/1212.6178}{\tt arXiv:1212.6178}.
\bibitem[{Tony~Hey and Tolle(2009)}]{Hey2009}
\bibinfo{editor}{Tony~Hey S.~T.}, \bibinfo{editor}{Tolle K.} (Eds.),
  \bibinfo{year}{2009}.
\newblock \bibinfo{title}{The Fourth Paradigm: Data-Intensive Scientific
  Discovery}.
\newblock \bibinfo{publisher}{Microsoft Research}.
\bibitem[{{van Diepen}(2015)}]{vanDiepen2015}
\bibinfo{author}{{van Diepen} G.~N.~J.}, \bibinfo{year}{2015}.
\newblock \bibinfo{title}{{Casacore Table Data System and its use in the
  MeasurementSet}}.
\newblock \bibinfo{journal}{Astronomy and Computing} \bibinfo{volume}{12},
  \bibinfo{pages}{174--180}.
\newblock \DOIprefix\doi{10.1016/j.ascom.2015.06.002}.
\bibitem[{{van Haarlem} et~al.(2013){van Haarlem}, {Wise}, {Gunst}, {Heald},
  {McKean}, {Hessels}, {de Bruyn}, {Nijboer}, {Swinbank}, {Fallows},
  {Brentjens}, {Nelles}, {Beck}, {Falcke}, {Fender}, {H{\"o}randel},
  {Koopmans}, {Mann}, {Miley}, {R{\"o}ttgering}, {Stappers}, {Wijers},
  {Zaroubi}, {van den Akker}, {Alexov}, {Anderson}, {Anderson}, {van Ardenne},
  {Arts}, {Asgekar}, {Avruch}, {Batejat}, {B{\"a}hren}, {Bell}, {Bell}, {van
  Bemmel}, {Bennema}, {Bentum}, {Bernardi}, {Best}, {B{\^i}rzan}, {Bonafede},
  {Boonstra}, {Braun}, {Bregman}, {Breitling}, {van de Brink}, {Broderick},
  {Broekema}, {Brouw}, {Br{\"u}ggen}, {Butcher}, {van Cappellen}, {Ciardi},
  {Coenen}, {Conway}, {Coolen}, {Corstanje}, {Damstra}, {Davies}, {Deller},
  {Dettmar}, {van Diepen}, {Dijkstra}, {Donker}, {Doorduin}, {Dromer}, {Drost},
  {van Duin}, {Eisl{\"o}ffel}, {van Enst}, {Ferrari}, {Frieswijk}, {Gankema},
  {Garrett}, {de Gasperin}, {Gerbers}, {de Geus}, {Grie{\ss}meier}, {Grit},
  {Gruppen}, {Hamaker}, {Hassall}, {Hoeft}, {Holties}, {Horneffer}, {van der
  Horst}, {van Houwelingen}, {Huijgen}, {Iacobelli}, {Intema}, {Jackson},
  {Jelic}, {de Jong}, {Juette}, {Kant}, {Karastergiou}, {Koers}, {Kollen},
  {Kondratiev}, {Kooistra}, {Koopman}, {Koster}, {Kuniyoshi}, {Kramer},
  {Kuper}, {Lambropoulos}, {Law}, {van Leeuwen}, {Lemaitre}, {Loose}, {Maat},
  {Macario}, {Markoff}, {Masters}, {McFadden}, {McKay-Bukowski}, {Meijering},
  {Meulman}, {Mevius}, {Middelberg}, {Millenaar}, {Miller-Jones}, {Mohan},
  {Mol}, {Morawietz}, {Morganti}, {Mulcahy}, {Mulder}, {Munk}, {Nieuwenhuis},
  {van Nieuwpoort}, {Noordam}, {Norden}, {Noutsos}, {Offringa}, {Olofsson},
  {Omar}, {Orr{\'u}}, {Overeem}, {Paas}, {Pandey-Pommier}, {Pandey}, {Pizzo},
  {Polatidis}, {Rafferty}, {Rawlings}, {Reich}, {de Reijer}, {Reitsma},
  {Renting}, {Riemers}, {Rol}, {Romein}, {Roosjen}, {Ruiter}, {Scaife}, {van
  der Schaaf}, {Scheers}, {Schellart}, {Schoenmakers}, {Schoonderbeek},
  {Serylak}, {Shulevski}, {Sluman}, {Smirnov}, {Sobey}, {Spreeuw}, {Steinmetz},
  {Sterks}, {Stiepel}, {Stuurwold}, {Tagger}, {Tang}, {Tasse}, {Thomas},
  {Thoudam}, {Toribio}, {van der Tol}, {Usov}, {van Veelen}, {van der Veen},
  {ter Veen}, {Verbiest}, {Vermeulen}, {Vermaas}, {Vocks}, {Vogt}, {de Vos},
  {van der Wal}, {van Weeren}, {Weggemans}, {Weltevrede}, {White}, {Wijnholds},
  {Wilhelmsson}, {Wucknitz}, {Yatawatta}, {Zarka}, {Zensus} and {van
  Zwieten}}]{LOFAR}
\bibinfo{author}{{van Haarlem} M.~P.} et~al., \bibinfo{year}{2013}.
\newblock \bibinfo{title}{{LOFAR: The LOw-Frequency ARray}}.
\newblock \bibinfo{journal}{\aap} \bibinfo{volume}{556}, \bibinfo{pages}{A2}.
\newblock \DOIprefix\doi{10.1051/0004-6361/201220873},
  \href{http://arxiv.org/abs/1305.3550}{\tt arXiv:1305.3550}.
\bibitem[{{van Weeren} et~al.(2016a){van Weeren}, {Brunetti}, {Br{\"u}ggen},
  {Andrade-Santos}, {Ogrean}, {Williams}, {R{\"o}ttgering}, {Dawson}, {Forman},
  {de Gasperin}, {Hardcastle}, {Jones}, {Miley}, {Rafferty}, {Rudnick},
  {Sabater}, {Sarazin}, {Shimwell}, {Bonafede}, {Best}, {B{\^i}rzan},
  {Cassano}, {Chy{\.z}y}, {Croston}, {Dijkema}, {En{\ss}lin}, {Ferrari},
  {Heald}, {Hoeft}, {Horellou}, {Jarvis}, {Kraft}, {Mevius}, {Intema},
  {Murray}, {Orr{\'u}}, {Pizzo}, {Sridhar}, {Simionescu}, {Stroe}, {van der
  Tol} and {White}}]{vanWeeren2016b}
\bibinfo{author}{{van Weeren} R.~J.} et~al., \bibinfo{year}{2016}a.
\newblock \bibinfo{title}{{LOFAR, VLA, and Chandra Observations of the
  Toothbrush Galaxy Cluster}}.
\newblock \bibinfo{journal}{\apj} \bibinfo{volume}{818}, \bibinfo{pages}{204}.
\newblock \DOIprefix\doi{10.3847/0004-637X/818/2/204},
  \href{http://arxiv.org/abs/1601.06029}{\tt arXiv:1601.06029}.
\bibitem[{{van Weeren} et~al.(2016b){van Weeren}, {Williams}, {Hardcastle},
  {Shimwell}, {Rafferty}, {Sabater}, {Heald}, {Sridhar}, {Dijkema}, {Brunetti},
  {Br{\"u}ggen}, {Andrade-Santos}, {Ogrean}, {R{\"o}ttgering}, {Dawson},
  {Forman}, {de Gasperin}, {Jones}, {Miley}, {Rudnick}, {Sarazin}, {Bonafede},
  {Best}, {B{\^i}rzan}, {Cassano}, {Chy{\.z}y}, {Croston}, {Ensslin},
  {Ferrari}, {Hoeft}, {Horellou}, {Jarvis}, {Kraft}, {Mevius}, {Intema},
  {Murray}, {Orr{\'u}}, {Pizzo}, {Simionescu}, {Stroe}, {van der Tol} and
  {White}}]{vanWeeren2016}
\bibinfo{author}{{van Weeren} R.~J.} et~al., \bibinfo{year}{2016}b.
\newblock \bibinfo{title}{{LOFAR Facet Calibration}}.
\newblock \bibinfo{journal}{\apjs} \bibinfo{volume}{223}, \bibinfo{pages}{2}.
\newblock \DOIprefix\doi{10.3847/0067-0049/223/1/2},
  \href{http://arxiv.org/abs/1601.05422}{\tt arXiv:1601.05422}.
\bibitem[{{Williams} et~al.(2016){Williams}, {van Weeren}, {R{\"o}ttgering},
  {Best}, {Dijkema}, {de Gasperin}, {Hardcastle}, {Heald}, {Prandoni},
  {Sabater}, {Shimwell}, {Tasse}, {van Bemmel}, {Br{\"u}ggen}, {Brunetti},
  {Conway}, {En{\ss}lin}, {Engels}, {Falcke}, {Ferrari}, {Haverkorn},
  {Jackson}, {Jarvis}, {Kapi{\'n}ska}, {Mahony}, {Miley}, {Morabito},
  {Morganti}, {Orr{\'u}}, {Retana-Montenegro}, {Sridhar}, {Toribio}, {White},
  {Wise} and {Zwart}}]{Williams2016}
\bibinfo{author}{{Williams} W.~L.} et~al., \bibinfo{year}{2016}.
\newblock \bibinfo{title}{{LOFAR 150-MHz observations of the Bo{\"o}tes field:
  catalogue and source counts}}.
\newblock \bibinfo{journal}{\mnras} \bibinfo{volume}{460},
  \bibinfo{pages}{2385--2412}.
\newblock \DOIprefix\doi{10.1093/mnras/stw1056},
  \href{http://arxiv.org/abs/1605.01531}{\tt arXiv:1605.01531}.

\end{thebibliography}

\end{document}